\documentclass[pra,aps,amsmath,amssymb,amsfonts,twocolumn,nofootinbib,longbibliography]{revtex4-1}
\usepackage{}
\usepackage{amssymb}
\usepackage{bm,mathrsfs}
\usepackage{graphicx}
\usepackage{epsfig}
\usepackage{amsmath,bbm}
\usepackage{amsfonts,amssymb}
\usepackage{times}
\usepackage{verbatim}
\usepackage[sort&compress]{natbib}
\usepackage{amsmath}
\usepackage{bm}
\usepackage{float}
\usepackage{textgreek}
\allowdisplaybreaks[4]
\usepackage[colorlinks,breaklinks,linkcolor=blue,anchorcolor=blue,citecolor=blue,urlcolor=magenta]{hyperref}

\usepackage{color}
\begin{document}
\title{Nonreciprocity and unidirectional invisibility in three optical modes with non-Markovian effects}

\author{H. Yi$^{1}$}

\author{T. Z. Luan$^{2}$}

\author{W. Y. Hu$^{1}$}

\author{Cheng Shang$^{3,4}$}
\email{Contact author: cheng.shang@riken.jp}

\author{Yan-Hui Zhou$^{5}$}
\email{Contact author: yanhuizhou@126.com}

\author{Zhi-Cheng Shi$^{6}$}

\author{H. Z. Shen$^{1}$}
\email{Contact author: shenhz458@nenu.edu.cn}

\makeatletter
\renewcommand\frontmatter@affiliationfont{\vspace{1mm} \small}
\makeatother

\affiliation{\textit{$^{1}$Center for Quantum Sciences and School of Physics, Northeast Normal University, Changchun 130024, China\\
$^{2}$School of Physics science and Technology, Shenyang Normal University, ShenYang 110034, China\\
$^{3}$Analytical quantum complexity RIKEN Hakubi Research Team,
RIKEN Center for Quantum Computing (RQC), Wako, Saitama 351-0198, Japan\\
$^{4}$Department of Physics, The University of Tokyo, 5-1-5 Kashiwanoha, Kashiwa, Chiba 277-8574, Japan\\
$^{5}$Quantum Information Research Center and Jiangxi Province Key Laboratory of Applied Optical Technology, Shangrao Normal University, Shangrao 334001, China\\
$^{6}$Department of Physics, Fuzhou University, Fuzhou 350116, China}}

\date{\today}

\begin{abstract}
In this work, we construct three coupled optical modes systems to obtain effective Hamiltonian mediated by coherent dissipative coupling during adiabatic elimination of large dissipation mode. We investigate the cooperative effect of coherent and dissipative photon-photon couplings in an open cavity system, which leads to nonreciprocity with a considerably large isolation ratio and flexible controllability. We discover unidirectional invisibility for electromagnetic wave propagation, which appears at the zero-damping condition (ZDC) for hybrid photon-photon modes and obtain transmission spectrum on the ZDC. We study the influences of the parameters on the nonreciprocal transmission of the system to capture the generic physics of the interference between coherent and dissipative couplings, which accurately reproduces the results of numerical simulation over a broad range of parameters. Moreover, we extend the study of nonreciprocal transmission with the Markovian approximation to the non-Markovian environments, which consist of a collection of oscillators (bosonic photonic modes) and give the adiabatic elimination method with non-Markovian effects. We illustrate that nonreciprocal transmission on ZDC exhibits a crossover from the non-Markovian to the Markovian regimes by controlling the environmental spectral width. This indicates a promising way to enhance or steer quantum nonreciprocal devices in optical cavities and provides potential applications for precision measurements and optical communications with non-Markovian effects.
\end{abstract}



\maketitle
\section{Introduction}\label{Sec1}
Reciprocity is a common phenomenon in nature. However, nonreciprocity has attracted significant attention from researchers
due to its indispensable applications in areas such as optical isolation \cite{Caloz0470012018,Bi758,Yu91},
quantum networks \cite{Lodahl4732009,Clark177901,YaoS318,Hong052302,Reiserer041003}, and optical information
processing \cite{Jalas5792013,Rohde052332,Li064083}. In recent years, optical nonreciprocity has been extensively explored in fields such as exotic topological
photonics \cite{Lu8212014,Khanikaev763, Hafezi1001}, realization of the chiral edge states \cite{Wang7722009,Barczyk574},
nonreciprocal photon blockades \cite{Huang1536012018,Li6302019,Shang115202,Sun043715,Liu063701, Jing033707,Xie053707,Gou043723},
the nonreciprocal slow light \cite{Jiang3672017,Mirza255152011,Tsakmakidis190501}, topological protection \cite{Hafezi9072011},
nonreciprocal phonon lasers \cite{Jiang0640372018}, and ground-state cooling \cite{Lai0115022020,Yang52507,Lai023860,Huang013526}. Generally, nonreciprocity is commonly got via Faraday
rotation \cite{Faraday1839C,Hogan2531953,Rowen13331953,Adam7212002,Kuanr202505}, which is difficult in integration and miniaturization. To deal with this challenge, researchers increasingly prefer magnetism-free nonreciprocal approaches, including optomechanical
interactions \cite{Bernier6042017,Peterson0310012017,Miri0640142017,Shen6572016}, nonlinear optics \cite{Fan447,Cao033901,Dimitrios359,Li2200267}, parametric time modulation \cite{Sounas7742017,Ranzani1030272014}, and atomic gases \cite{Wang093901,Ramezani043901,Yang2200574,Gou070402}. Additionally, dissipative coupling can be induced by reservoir engineering \cite{Hao250113140,Lidissipative2025,Liu032403,Hou072401,Yu115012,Lu221101,Li233101,Dong2300212,Hu2100534,Liu2200660,Peng043527,Zhang012410,Hong052201,Rao021003,Zhao044074,Rao064404,Nair224401,Yang054413,Rao1933}, which has been employed to achieve nonreciprocal effects \cite{Metelmann021025} by balancing coherent and dissipative interactions between different modes. To date, the reservoir engineering has found applications in a wide range of nonreciprocal phenomena, including but not limited to nonreciprocal transmission \cite{Fang4652017,Goulon43852000,Shang2019nonreciprocity,Munir116019,Ming107915,Shangcoupling2023,Biehs035423,Chen042403,Kim063904,Luantouter2025,Shi132403,Qian192401,Dong25726,Zhao014035,Han4730}, nonreciprocal entanglement \cite{Ren06775,Zhan7032}, nonreciprocal blockade \cite{Cong108023,Lu053029,Deng108018} and one-way steering \cite{Ye36907,Yang044056,Guan102}. In practice, achieving large and flexible controllable nonreciprocity remains a significant issue\cite{Caloz0470012018,Shi3882015}.

In fact, all quantum systems  \cite{Breuer2002,Weiss2012} interact with the external environments \cite{Caruso12032014,Chengopen2024,Gardiner2000,Li0621242010,Lu1082023open,Franco13450532013}, which means those are open systems, so the study of open systems is quite important in practical applications. Markovian approximation is reasonable as a simplified method \cite{Breuer2002,Weiss2012} when the coupling strength between the system and the environment \cite{Caruso12032014,Breue0453232004,Li56142022,Ferraro0421122009,Shen315042019,Xu0441052010,Shen28522018,He0121082011}is weak,. However, when the interact strengths between the systems and the environments are strong, and the characteristic time scale of the system is equivalent to the relaxation time scale of the environment, it is necessary to consider the non-Markovian effect \cite{Reuther0621232012}. In the fields of quantum state engineering, quantum channel capacity and quantum control \cite{Bylicka57202014,Xue0523042012}, non-Markovian effects caused by environment have been observed in many systems, including coupled cavities \cite{Link0203482022}, photonic crystals \cite{Burgess0436032012}, color noises \cite{Costa-Filho0521262017}, cavities coupled to waveguides \cite{Chang0521052010,Longhi0638262006,Tan0321022011,LuTQB2024, Zheng0340352020,Vega0150012017}, and successfully realized experimentally \cite{Liu9312011,Xiong0321012019,Cialdi0521042019,Zhang0638532016, Khurana0221072019,Madsen2336012011,Guo2304012021,Li1405012022,Uriri0521072020, Anderson32021993,Liu0622082020,Fanchini2104022014,Haseli0521182014,Goswami0224322021,Debiossac2006012022}. Unlike the Markovian effect, the excitation return between the systems and the environments can characterize the non-Markovian effects on the systems dynamics \cite{Breuer0210022016,Breuer2104012009, Breuer0621082012,ShenNHgain2025,Breuer0421082015,Xin0537062022}, which can be quantified by various  measures of non-Markovianity \cite{Lorenzo0201022013,Rivas0504032010,Luo044101201, Wolf1504022008,Lu0421032010,Maniscalco1204042014,Hou0621152011,Hou0121012012}.

In another way, open quantum systems are usually described by typical non-Hermitian Hamiltonians, where the eigenvalues are usually complex. A distinct class of non-Hermitian physical systems respecting parity-time ($\mathcal{PT}$) symmetry is characterized by real energy spectra in the $\mathcal{PT}$ symmetric phase \cite{Bender52431998, Mostafazadeh2052002,Mostafazadeh28142002,Mostafazadeh39442002}. The phase transition occurs in the parameter space \cite{Bender2772005,Bender9472007, Grigoryan0644262020,Konotop0350022016} by varying the parameters, where their complex eigenspectra exhibit exotic degeneracies at the branch-point singularities \cite{Grigoryan0244062018,Grigoryan2244082019,Heiss4440162012}, i.e., the exceptional point (EP). Specially, many interesting phenomena have been discovered in the vicinity of EP, such as
the enhanced mechanical cooling \cite{Lu0637082021}, spontaneous emission \cite{Lin1074022016}, lowering of chaos threshold power \cite{Jing2536012015}, unidirectional invisibility \cite{Jing2536012015,Feng1082013,Lin2139012011}, and lasing \cite{Feng9722014,Hodaei9752014,Liertzer1739012012}. Especially, the dispersion readout of the system is enhanced near the EP \cite{Liu1108022016,Li0538372015,Wiersig2039012014}. Moreover, there is another special class of non-Hermitian systems, which satisfy \{$\mathcal{PT}$,$H_{a\mathcal{PT}}$\}= 0. Unlike $\mathcal{PT}$-symmetric systems which usually requires gain mediums \cite{Rotter7832019}, the anti-$\mathcal{PT}$(a$\mathcal{PT}$)
symmetry based on pure loss \cite{Ge0538102013,Peng11392016,Huang031503,Yang147202,Luo173602,Qi063520,Hu134110,Ding010204,Wang013131,Zhang053901,Peng043527,Yang033504,Fang033022,Peng033507,Yang053845,Mukhopadhyay064405,Xu012218,Wang013131,Zhangantipt2024,Nair180401} has also attracted much attention due to their intriguing properties such
as energy-difference conserving dynamics \cite{Park0836012021,Choig21822018} and shorter-length chiral mode switch \cite{Feng2736012022}. In experiments, a$\mathcal{PT}$ symmetry and breaking have been observed in a wide range of systems, such as radiative plasmonics \cite{Yang76782022}, thermal or cold atoms \cite{Peng11392016,Ding010204,Jiang1936042019,Cao0304012020,He0437122022}, electrical circuits \cite{Choig21822018,Stegmaier2153022021}, magnonic systems \cite{Yang147202,Zhao0140532020}, optical waveguides or microcavities \cite{Zhang053901,Wang0440502020,Zhang882019,Fan30352020,Bergman4862021}, and diffusive systems \cite{Li1702019,Cao482020}.

Such recipes as those above lead us to inquire into these issues: (i)~How can we construct the nonreciprocal model in non-Markovian environments? (ii)~Can transmission efficiency be converted by generalizing nonreciprocal transmission from Markovian systems to non-Markovian ones? (iii)~How can non-Markovian effects affect nonreciprocal transmission efficiency?

To address these problems, we propose a scheme to realize nonreciprocal transmission in non-Markovian environments by applying an optical system, which consists of three interacting optical cavities coupling with three environments. The effective Hamiltonian can be constructed via adiabatic elimination by using a cavity with large dissipation. Consequently, the system becomes a two-mode system mediated by coherent coupling and dissipative coupling. We demonstrate that the time-reversal symmetry can be broken through the interference of coherent and dissipative photon coupling, where the system can realize nonreciprocal transmission. By exploring the influences of different parameters on nonreciprocal transmission, we show that the scheme of nonreciprocity has flexible controllability, which enables a high isolation ratio accompanied by low insertion loss. Moreover, we derive the adiabatic elimination procedure including non-Markovian effects and point out nonreciprocal transmission phenomena still exist in non-Markovian environments. As the environmental spectrum widths increase, the results with the non-Markovian environments can return to the Markovian ones in the wideband limit.

This paper is organized as follows. In Sec.~\ref{Sec2} and Sec.~\ref{Sec3}, we derive the transmission coefficient of the introduced system and study the nonreciprocal transmission phenomenon of the system under the Markovian approximation. In Sec.~\ref{Sec4} - Sec.~\ref{Sec5}, we obtain the effective Hamiltonian of the system in non-Markovian environments by adiabatic elimination. The exact non-Markovian input-output relations can return to the Markovian input-output ones in the wideband limit. Through numerical simulation, we get the non-Markovian transmission spectrum of the system and explore the relationship between spectral widths and transmission efficiency. The main conclusions are summarized in Sec.~\ref{Sec6}.
\begin{figure}[t]
\centerline{
\hspace{0mm}
\includegraphics[width=7.8cm, height=5.6cm, clip]{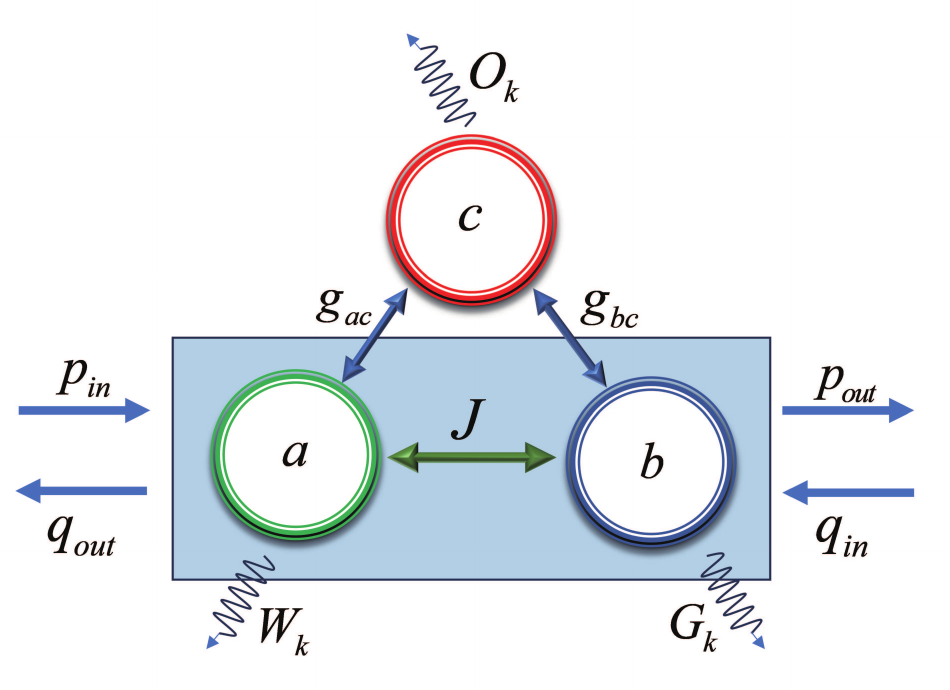}}
\vspace{-0.1cm}
\caption{A nonreciprocal transmission device in non-Markovian environments, which may be realized in superconducting circuits \cite{Zhong1539022019,Qin20005692021,
Li0335052021,Carlo59542021,Carlo60212022,Jiang2011062023,Zhong0132202021,
Zhong52422019,Soleymani5992022,Tang66602023,Jia2932023,Stalhammar2011042021,Chen0315012022,Zhong0140702020}. The coupled cavities $a$ and $b$ (the interacting strength $J$) interact with cavity $c$ via the strengths $g_{ac}$ and $g_{bc}$. The three optical photon modes couple to three non-Markovian environments with eigenfrequencies $x_{k}, n_k$, and $y_k$ (annihilation operators $\hat{d}_k$, $\hat{e}_k$, and $\hat{f}_k$), where ${W}_k$, ${G}_k$ and ${O}_k$ denote coupling coefficients between three cavities and three non-Markovian environments.}\label{model}
\end{figure}

\section{THE MODEL AND METHODS}\label{Sec2}
We propose a scheme to obtain nonreciprocity and unidirectional invisibility  in non-Markovian environments based on coherent interaction between different optical photons in Fig.~\ref{model}. It is a quantum system consisting of three bosonic systems (frequencies $\omega_{a}, \omega_b$ and $\omega_c$) with three non-Markovian environments (eigenfrequencies $x_{k}, n_k$ and $y_k$), where three optical cavities interact with three different non-Markovian environments (coupling strengths $W_k, G_k$ and $O_k$), which are sketched in Fig.~\ref{model}. In Sec.~\ref{Sec2} - Sec.~\ref{Sec3}, we explore Markovian cases, while Sec.~\ref{Sec4} - Sec.~\ref{Sec5} consider the influences of the non-Markovian effects on the nonreciprocity. The Hamiltonian of the system is written as follows \cite{Zhao0140532020}:
\begin{equation}
\begin{aligned}
\hat{H}=~&\omega_a\hat{a}^\dagger\hat{a}+\omega_b\hat{b}^\dagger\hat{b}+\omega_c\hat{c}^\dagger\hat{c}\\
&+(g_{ac}\hat{a}^\dagger\hat{c}+g_{bc}e^{i\phi}\hat{b}^\dagger\hat{c}+J\hat{a}^\dagger\hat{b}+\text{H.c.}),\label{H11}
\end{aligned}
\end{equation}where $\hat{a}(\hat{a}^\dagger),\hat{b}(\hat{b}^\dagger)$ and $\hat{c}(\hat{c}^\dagger)$ are the optical modes annihilation (creation) operators. Herein, $J$, $g_{ac}$ and $g_{bc}$ correspond to optical modes coupling strengths, respectively. The angle $\phi$ represents the physical phase of the system. With the operators expection values defined by $a\equiv\left\langle\hat{a}\right\rangle$, $b \equiv \langle \hat{b} \rangle$ and $c\equiv\left\langle\hat{c}\right\rangle$, the corresponding Heisenberg-Langevin equations with Eq.~({\ref{H11}) for envionments initialization in the vacuum states are given by
\begin{eqnarray}
\!\!\dot{a}(t)\!&=&\!-(i\omega_a+\Gamma_1/2)a(t)-ig_{ac}c(t)-iJb(t),
\label{dota1}\\
\!\!\dot{b}(t)\!&=&\!-(i\omega_b+\Gamma_2/2)b(t)-ig_{bc}e^{i\phi}c(t)-iJa(t),
\label{dotb1}\\
\!\!\dot{c}(t)\!&=&\!-(i\omega_c+\Gamma_3/2)c(t)-ig_{ac}a(t)-ig_{bc}e^{-i\phi}b(t).
\label{dotc1}
\end{eqnarray}
Introducing the slowly varying amplitudes $A(t), B(t)$ and $C(t)$ with
\begin{eqnarray}
a(t)&=&A(t)e^{-i\omega_{a}t},
\label{5}\\
b(t)&=&B(t)e^{-i\omega_{b}t},
\label{6}\\
c(t)&=&C(t)e^{-i\omega_{c}t},
\label{7}
\end{eqnarray}
According to Eqs.~(\ref{5})-(\ref{7}), we can derive the equations of motion for slowly varying amplitudes as
\begin{align}
\!\dot{A}(t) \!&=\!  -\frac{\Gamma_{1}}{2}A(t) \!- ig_{ac}C(t)e^{i\Delta_{ac}t} \!- iJB(t)e^{i\Delta_{ab}t},\! \label{dotA} \\
\!\dot{B}(t) \!&=\! -\frac{\Gamma_{2}}{2}B(t) \!\!- ig_{bc}e^{i\phi}C(t)e^{i\Delta_{bc}t} \!- iJA(t)e^{-i\Delta_{ab}t},\! \label{dotB} \\
\!\dot{C}(t) \!&=\! -\frac{\Gamma_{3}}{2}C(t) \!- ig_{ac}A(t)e^{-i\Delta_{ac}t} \!- ig_{bc}e^{-i\phi}B(t)e^{-i\Delta_{bc}t}, \nonumber
\end{align}
where $\Delta_{ac}=\omega_a-\omega_c$, $\Delta_{bc}=\omega_b-\omega_c$ and $\Delta_{ab}=\omega_a-\omega_b$, which are the frequency detunings between three optical cavities. By solving $\dot C\left( t \right)$, we can obtain the solution of $C(t)$ as
\begin{eqnarray}
\begin{aligned}
C(t)=&-ig_{ac}\int_0^{t}dt'A(t')e^{-i\Delta_{ac}t'}e^{-\Gamma_3/2(t-t')}\\
&-ig_{bc}e^{-i\phi}\int_0^{t}dt'B(t')e^{-i\Delta_{bc}t'}e^{-\Gamma_3/2(t-t')}.
\label{C2}
\end{aligned}
\end{eqnarray}
If the dissipation rate $\Gamma_3$ of the optical mode $c$ is much larger than $\Gamma_1, \Gamma_2$ ($\Gamma_3\gg\Gamma_1, \Gamma_2$), the amplitude changes of mode $a$ and mode $b$ are small within the range of the integration of the cavity mode $c$. The approximations $A(t') = A(t)$ and $B(t') = B(t)$ are reasonable, so that the equation of motion for C can be simplified to
\begin{eqnarray}
\begin{aligned}
C(t)=&-\frac{ig_{ac}}{\Gamma_3/2-i\Delta_{ac}}A(t)e^{-i\Delta_{ac}t}\\
&-\frac{ig_{bc}e^{-i\phi}}{\Gamma_3/2-i\Delta_{bc}}B(t)e^{-i\Delta_{bc}t}.
\label{C3}
\end{aligned}
\end{eqnarray}
Substituting Eq.~(\ref{C3}) into Eqs.~\eqref{dotA} and~\eqref{dotB} can adiabatically eliminate the mode $c$. This process produces a set of simplified equations of motion that describe only the evolution of modes $a$ and $b$
\begin{align}
\!\!\dot{A}(t) \!&=\! -{\Gamma_{1}}/{2}*A(t) - \frac{g_{ac}^{2}}{\Gamma_{3}/2 - i\Delta_{ac}} A(t) \nonumber \\
&\quad - \frac{g_{ac}g_{bc}e^{-i\phi}}{\Gamma_{3}/2 - i\Delta_{bc}} B(t)e^{i\Delta_{ab}t} - iJB(t)e^{i\Delta_{ab} t}, \label{dotAe} \\
\!\!\dot{B}(t) \!&=\! -{\Gamma_{2}}/{2}*B(t) - \frac{g_{bc}^{2}}{\Gamma_{3}/2 - i\Delta_{bc}} B(t) \nonumber \\
&\quad - \frac{g_{ac}g_{bc}e^{i\phi}}{\Gamma_{3}/2 - i\Delta_{ac}} A(t)e^{-i\Delta_{ab}t} - iJA(t)e^{-i\Delta_{ab} t}. \label{dotBe}
\end{align}
Through Eqs.~(\ref{dotAe}) and (\ref{dotBe}), we have
\begin{align}
\dot{a}(t) &= \left[-i\omega_a - \left({\Gamma_{1}}/{2} + \frac{g_{ac}^{2}}{\Gamma_{3}/2 - i\Delta_{ac}}\right)\right] a(t) \nonumber \\
&\quad - \frac{g_{ac}g_{bc}e^{-i\phi}}{\Gamma_{3}/2 - i\Delta_{bc}} b(t) - iJb(t), \label{dotCx} \\
\dot{b}(t) &= \left[-i\omega_b - \left({\Gamma_{2}}/{2} + \frac{g_{bc}^{2}}{\Gamma_{3}/2 - i\Delta_{bc}}\right)\right] b(t) \nonumber \\
&\quad - \frac{g_{ac}g_{bc}e^{i\phi}}{\Gamma_{3}/2 - i\Delta_{ac}} a(t) - iJa(t). \label{dotCe}
\end{align}
The effective Hamiltonian of the system corresponding Eqs.~(\ref{dotCx}) and (\ref{dotCe}) can be written as
\begin{eqnarray}
\begin{matrix}
\hat{H}=\left[
\begin{array}{cc}
 \omega_a-i {K_2} & J-i K_4 \\
 J-i {K_3} & \omega_b-i {K_1} \\
\end{array}
\right],
\end{matrix}\label{Heff}
\end{eqnarray}
where
\begin{align}
{K_1} &= \frac{{g_{bc}^2}}{{(\frac{{{\Gamma _3}}}{2} - i{\Delta _{bc}})}},{K_2} = \frac{{g_{ac}^2}}{{(\frac{{{\Gamma _3}}}{2} - i{\Delta _{ac}})}},\nonumber\\
{K_3} &= \frac{{{g_{ac}}{g_{bc}}{e^{i\phi }}}}{{(\frac{{{\Gamma _3}}}{2} - i{\Delta _{ac}})}},{K_4} = \frac{{{g_{ac}}{g_{bc}}{e^{ - i\phi }}}}{{(\frac{{{\Gamma _3}}}{2} - i{\Delta _{bc}})}}.\label{K1-K4}
\end{align}
The $K_3$ and $K_4$ are defined as the effective dissipative coupling strength.
According to Eq.~\eqref{Heff}, we obtain two eigenvalues
\begin{align}
&\omega_{\pm}=\frac{1}{2}\left(\omega_a+\omega_b-iK_1-iK_2\right)\label{eig}\\
&\pm\frac{1}{2}\sqrt{4(J-iK_3)(J-iK_4)-(K_2-K_1+i(\omega_a-\omega_b))^2},\nonumber
\end{align}
which correspond to two hybridized modes. The real parts of $\omega_{\pm}$ denote the dispersions of two hybridized
modes, while the imaginary parts correspond to their damping rates. In particular, When the appropriate parameters are taken, their damping rates Im$[\omega_{\pm}]$ may go to zero, in which is defined as ZDC. In Fig.~\ref{tu2}(b), we plot both the real and imaginary parts of $\omega_\pm$ as a function of the detuning $\Delta$, where the black solid line corresponds to $\omega_+$, while the dotted red line represents $\omega_-$. These are calculated under condition $\phi = 0$. In contrast to that we plot both the real and imaginary parts of $\omega_\pm$ as a function of the detuning $\Delta$ (see Fig.~\ref{tu2}(d)), where the calculation is performed for $\phi = \pi$. Moreover, the quantum Langevin equations for the system are given by
\begin{align}
&\dot{a}(t) = \left[-i\omega_a - \left({\Gamma_{1}}/{2} + \frac{g_{ac}^{2}}{\Gamma_{3}/2 - i\Delta_{ac}}\right)\right] a(t) \label{dotCxa} \\
&- \frac{g_{ac}g_{bc}e^{-i\phi}}{\Gamma_{3}/2 - i\Delta_{bc}} b(t) - iJb(t)+\sqrt{\Gamma_1}p_{in}(t)+\sqrt{\Gamma_1}q_{in}(t), \nonumber\\
&\dot{b}(t) = \left[-i\omega_b - \left({\Gamma_{2}}/{2} + \frac{g_{bc}^{2}}{\Gamma_{3}/2 - i\Delta_{bc}}\right)\right] b(t) \label{dotCeb} \\
&- \frac{g_{ac}g_{bc}e^{i\phi}}{\Gamma_{3}/2 - i\Delta_{ac}} a(t) - iJa(t)+\sqrt{\Gamma_2}p_{in}(t)+\sqrt{\Gamma_2}q_{in}(t). \nonumber
\end{align}
Solving Eqs.~\eqref{dotCxa} and~\eqref{dotCeb} in the frequency regime obtains the steady-state solution
\begin{figure}[t]
\centerline{
\includegraphics[width=8.6cm, height=6.8cm, clip]{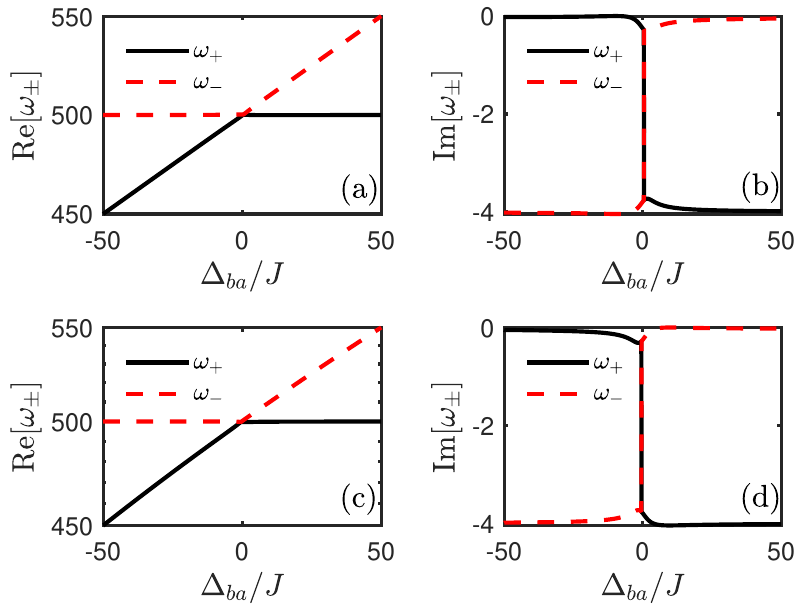}}
\vspace{-0.3cm}
\caption{The real and imaginary parts of $\omega_{\pm}$ are plotted as a function of the detuning $\Delta_{ba}=\omega_b-\omega_a$ by solving the real and imaginary parts of the eigenvalues according Eq.~(\ref{eig}). (a),(b) $\phi=0$. (c),(d) $\phi=\pi$. Black solid and red dotted lines respectively indicate the absolute values of $\omega_+$ and $\omega_-$. The parameters chosen are $\omega_a=500J$, $\omega_c=500J$, $\Gamma_1=10J$, $\Gamma_2=0.1J$, $\Gamma_3=100J$, $g_{ac}=20J$ and $g_{bc}=2J$.}\label{tu2}
\end{figure}
\begin{align}
\!&0 = \left[i(\omega - \omega_a) - \left({\Gamma_{1}}/{2} + \frac{g_{ac}^{2}}{\Gamma_{3}/2 - i\Delta_{ac}}\right)\right] a(\omega) \label{dotCxp} \\
\!\!&- \frac{g_{ac}g_{bc}e^{-i\phi}}{\Gamma_{3}/2 - i\Delta_{bc}} b(\omega) - iJ b(\omega)+\!\sqrt{\Gamma_1}p_{in}(\omega)+\!\sqrt{\Gamma_1}q_{in}(\omega),\nonumber\\
\!&0 = \left[i(\omega - \omega_b) - \left({\Gamma_{2}}/{2} + \frac{g_{bc}^{2}}{\Gamma_{3}/2 - i\Delta_{bc}}\right)\right] b(\omega) \label{dotCep} \\
\!\!&- \frac{g_{ac}g_{bc}e^{i\phi}}{\Gamma_{3}/2 - i\Delta_{ac}} a(\omega) - iJ a(\omega)+\!\sqrt{\Gamma_2}p_{in}(\omega)+\!\sqrt{\Gamma_2}q_{in}(\omega).\nonumber
\end{align}
Due to $\Gamma_1\gg\Gamma_2$, we ignore the $\sqrt{\Gamma_2}p_{in}(\omega)$ and $\sqrt{\Gamma_2}q_{in}(\omega)$. Thus, we obtain the input-output relationship for the two-sided cavity
\begin{eqnarray}
p_{out}(\omega)-p_{in}(\omega)&\approx-\sqrt{\Gamma_1}a(\omega)\label{Main},\\
q_{out}(\omega)-q_{in}(\omega)&\approx-\sqrt{\Gamma_1}a(\omega),
\label{Mbin}
\end{eqnarray}
we obtain
\begin{eqnarray}
\begin{aligned}
\mathbf{V_{out}(\omega)}=\mathbf{S^M(\omega)}\mathbf{V_{in}(\omega)},
\end{aligned}
\end{eqnarray}
or scattering matrix of the system
\begin{eqnarray}
\begin{matrix}
\mathbf{S^{M}(\omega)}=\left[
\begin{array}{cc}
 S_{12}^{M} & S_{11}^{M} \\
 S_{22}^{M} & S_{21}^{M} \\
\end{array}
\right],
\end{matrix}
\end{eqnarray}
where $\mathbf{V_ {out}(\omega)} = [p_{out}(\omega),q_{out}(\omega)]^T$ and $\mathbf{V_{in}(\omega)}\!=\![p_{in}(\omega),q_{in}(\omega)]^T$. Here, $S_{12,(21)}^{M}$ describes the photon transmission of the signal from left (right) to right (left) in the system. Thus, the transmission coefficient of the system can be obtained by
\begin{eqnarray}
\begin{aligned}
S_{12}^{M}=&1-\frac{\Gamma_{1}N}{-(iJ+K_3)(iJ+K_4^{\phi=0})+MN},\\
S_{21}^{M}=&1-\frac{\Gamma_{1}N}{-(iJ+K_3)(iJ+K_4^{\phi=\pi})+MN},\label{S2}
\end{aligned}
\end{eqnarray}
where $M={\Gamma_1}/{2}+K_2-i(\omega-\omega_a)$, $N={\Gamma_{2}}/{2}+K_1-i(\omega-\omega_b)$, and $\phi$ in Eq.~(\ref{S2}) takes the value of $0$ for $S_{12}^{NM}$ and $\pi$ for $S_{21}^{NM}$, respectively.

\section{DISCUSSION OF TRANSMISSION CHARACTERISTICS}\label{Sec3}

\begin{figure}[t]
\centerline{
\includegraphics[width=8.6cm, height=6.8cm, clip]{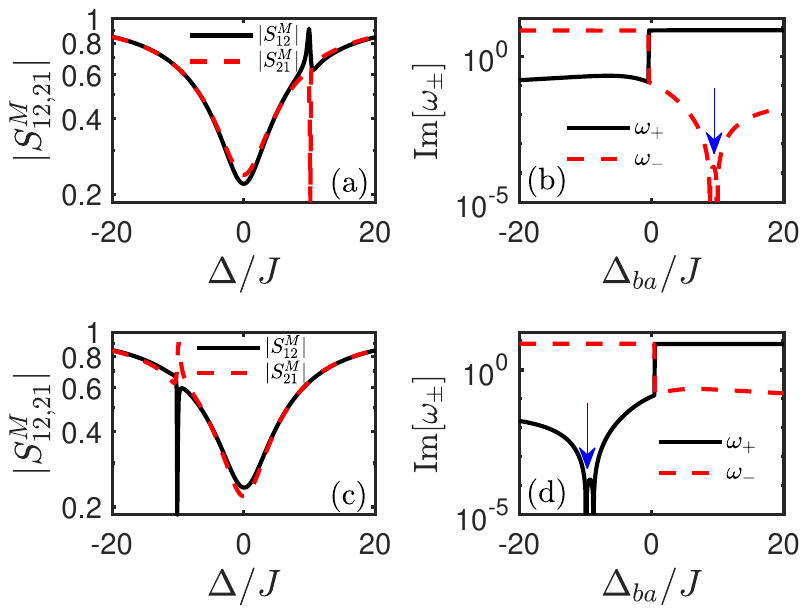}}
\vspace{-0.3cm}
\caption{Transmission spectrum $|S_{21,12}^{M}|$ in Eq.~(\ref{S2}) is plotted as a function of $\Delta=\omega-\omega_a$ for different direction. (a) Transmission spectrum of the system with $\omega_b-\omega_a=10J$. (b) The logarithm of the imaginary part of the system eigenvalues with $\phi=\pi$. (c) Transmission spectrum of the system with $\omega_b-\omega_a=-10J$. (d) The logarithm of the imaginary part of the system eigenvalues with $\phi=0$. The red dashed line and black solid line respectively denote different eigenvalues in Eq.~(\ref{eig}), while the ZDC on each curve is labelled by blue arrows. The other parameters are the same as in Fig. \ref{tu2}.}\label{tu3}
\end{figure}

Nonreciprocity means that electromagnetic wave can being transmitted in one direction but is suppressed in the other direction, while unidirectional invisibility means the electromagnetic wave is completely blocked when propagating in one direction. Now we demonstrate that the system transmission spectrum satisfies this requirement at the ZDC point. According to Eq.~(\ref{S2}), we plot the transmission spectra of the system when the light field is transmitted from different directions in Fig.~\ref{tu3}(a,c), where the black soild line $|S_{12}^{M}|$ represents the input of the light field from the left side, while the red line $|S_{21}^M|$ denotes the input from the right side. In Fig.~\ref{tu3}(a), besides the normal transmission spectrum, we observe a sharp regime characterized by nonreciprocal transmission located approximately at $\Delta = 10J$. At this point, the system shows left input enhancement while right input is suppressed. That is to say, electromagnetic wave transmission from the right to the left is blocked at particular frequency, which is the nonreciprocity of optical transmission. Clearly, this direction dependent nonreciprocal behavior can be attributed to the interference of both coherent and dissipative coupling, where the reason will be discussed below. In order to make the ZDC more clearly displayed, we plot the logarithmic of the imaginary part of the eigenvalues as a function of $\Delta$, in which ZDC points are marked by blue arrows in Fig.~\ref{tu3}(b,d). We find the logarithmic figure of the imaginary part reaches $10^{-5}$, which means the imaginary part approaches zero, precisely corresponding to the nonreciprocity transmission. All the above results are calculated by the parameters $\Gamma_1= 10J$, $g_{ac}=20J$, $g_{bc}=2J$ and $\Gamma_2= 0.1J$.

\begin{figure}[t]
\centerline{
\includegraphics[width=8.2cm, height=5.8cm, clip]{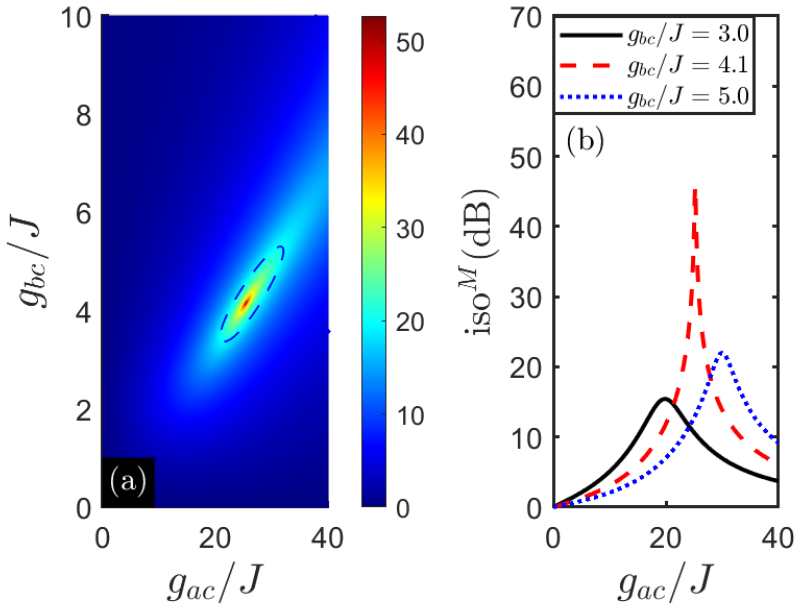}}
\vspace{-0.3cm}
\caption{The influences of coupling strength on isolation degree. (a) Isolation ratio in Eq.~(\ref{S2}) is plotted as a function of the coupling strength $g_{ac}$ and $g_{bc}$. The blue dashed line is the contour line of 20dB of isolation ratio. (b) Isolation ratio in Eq.~(\ref{S2}) near the side isolation peak is plotted as a function of the coupling strength $g_{ac}$ with different $g_{bc}$. The parameters chosen are $\omega_b=510J$, $\Delta=10J$ and the other parameters are the same as in Fig. \ref{tu3}.}\label{tu4}
\end{figure}

As a crucial performance metric for nonreciprocal systems, the isolation ratio directly reflects the degree of nonreciprocity, with greater isolation indicating better performance. Therefore, we study the nonreciprocity with different combinations of parameters. With Eq.~($\ref{S2})$, we can obtain analytical expressions for transmission coefficient and the difference between the $|S_{12}^{M}|$ and $|S_{21}^{M}|$ transmission amplitudes extracted in the decibel scale (defined as $20*\mathrm{log}_{10}|S_{12}^{M}/{S_{21}^{M}}|$), which takes its absolute value as the isolation ratio. Next, we discuss the case of the isolation ratio with different parameters. For the situation discussed in Fig.~\ref{tu4}, the isolation ratio as a function of the coupling strengths $g_{ac}$ and $g_{bc}$ is plotted, which is calculated by fixing the decays $\Gamma_1= 10J$ and $\Gamma_2= 0.1J$. In Fig.~\ref{tu4}(a), the isolation ratio above 50dB can be obtained, where the blue dashed line is the contour line of 20dB of isolation ratio. We can see that the isolation ratio gradually varies with the changes of the coupling strengths $g_{ac}$ and $g_{bc}$ in which reaching the maximum value at $g_{ac} = 25J$ and $g_{bc} = 4J$. In this case, we show more details near the peak of the isolation ratio in Fig.~\ref{tu4}(b). In Fig.~\ref{tu4}(b), the isolation ratio reaches the maximum as shown by the red dotted line. Moreover, we find the isolation ratio is very sensitive to parameter changes especially $g_{bc}$, in which the maximum value of isolation ratio only reaches 20dB when $g_{bc}$ takes other values, as shown black solid line and blue dotted line in the Fig.~\ref{tu4}(b). Therefore, we can flexibly control the isolation ratio by controlling the coupling strength.
\begin{figure}[t]
\centerline{
\includegraphics[width=8.2cm, height=5.8cm, clip]{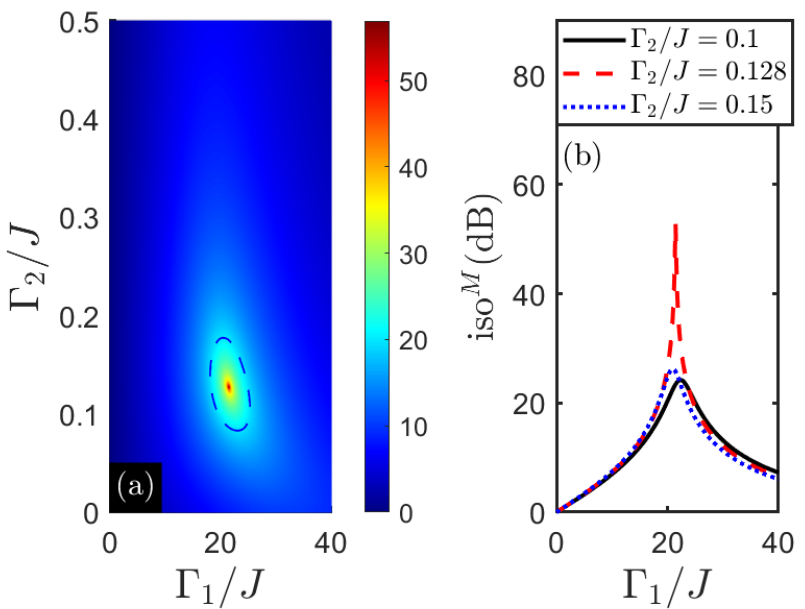}}
\vspace{-0.3cm}
\caption{The influences of external decay on isolation ratio according to Eq.~(\ref{S2}). (a) Isolation ratio is plotted as a function of the decay $\Gamma_1$ and the decay $\Gamma_2$. The blue dashed line is the contour line of 20dB of isolation ratio. (b) Isolation ratio near the side isolation peak is plotted as a function of the decay $\Gamma_1$ and the decay $\Gamma_2$. The parameters chosen are $\omega_b=510J$, $\Delta=10J$ and the other parameters are the same as in Fig. \ref{tu3}.}\label{tu5}
\end{figure}
\begin{figure}[t]
\centerline{
\includegraphics[width=8.4cm, height=5.9cm, clip]{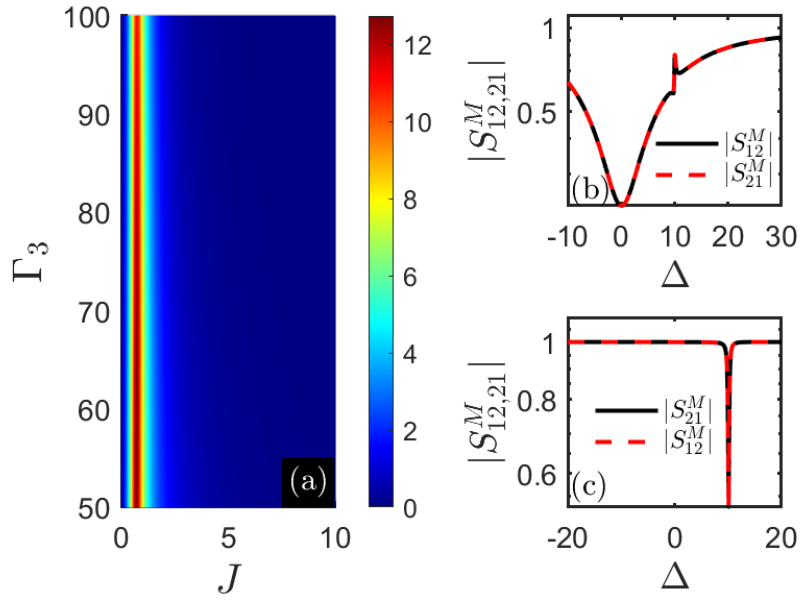}}
\vspace{-0.3cm}
\caption{Exploring the cause of nonreciprocity of systems. (a) Isolation ratio is plotted by using Eq.~(\ref{S2}) as a function of the decay $\Gamma_3$ and coupling strength $J$. (b)(c) Transmission spectrum in Eq.~(\ref{S2}) is plotted as a function of the detuning $\Delta$ with the coupling strength $J$ = 0 and $g_{ac}$ = 0. The parameters chosen are $\omega_b=510J$, $\Delta=10J$ and the other parameters are the same as in Fig. \ref{tu3}.}\label{tu6}
\end{figure}

Since $\Gamma_1$ and $\Gamma_2$ are external parameters, we can more conveniently alter the nonreciprocal transmission properties of the system by tuning them. Now, we study the nonreciprocity with different dampings. In Fig.~\ref{tu5}, we plot the isolation ratio as a function of $\Gamma_1$ and $\Gamma_2$ by fixing $g_{ac} = 20J$ and $g_{ab} = 2J$. In Fig.~\ref{tu5}(a), the blue dashed line is the contour line of 20dB of isolation ratio. It can be seen that the tunable range of decay is slightly larger than that of coupling strength with the same isolation level. In Fig.~\ref{tu5}(b), we plot isolation ratio as a function of $\Gamma_1$ with different values of $\Gamma_2$, in which the isolation ratio reaches its maximum value of 50dB when $\Gamma_2$ is equal to $0.128J$ as shown by the red solid line in Fig.~\ref{tu5}(b).

In Fig.~\ref{tu6}, we analyze the effects of parameters $\Gamma_3$ and $J$ to investigate the cause of nonreciprocity. We observe that the value of $\Gamma_3$ has little influences on the system isolation ratio, this is because the $c$ optical mode is adiabatically eliminated in our system with $\Gamma_3\gg\Gamma_1,~\Gamma_2$. Therefore, the value of $\Gamma_3$ varies significantly, while the isolation ratio changes very little. Moreover, in Fig.~\ref{tu6}(b,c), we plot the transmission spectrum, where the coupling strength $J$ or $g_{ac}$ is fixed to zero (we only consider $g_{ac}$ here since $g_{ac}$ and $g_{bc}$ are equivalent). We can see transmission spectra are completely consistent in different directions, that is to say, the system exhibits no nonreciprocal characteristics. It's obvious, according to Eq.~($\ref{S2})$, the term proportional to $J*g_{ac}$ arises due to the interference of coherent and dissipative couplings. It is precisely that this term causes the nonreciprocal behavior and becomes zero when $J$ or $g_{ac}$ is zero, therefore the nonreciprocity disappears.
\begin{figure}[t]
\centerline{
\includegraphics[width=8cm, height=5.8cm, clip]{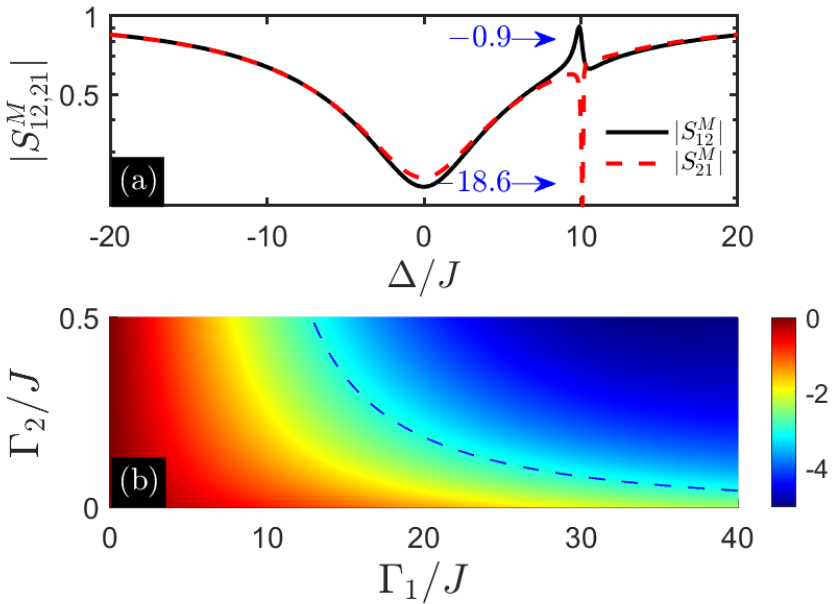}}
\vspace{-0.3cm}
\caption{The isolation and insertion loss calculated by Eq.~(\ref{S2}) of the system are studied under different parameters. (a) Transmission spectrum is plotted as a function of the detuning $\Delta$ with the isolation ratio 18.6dB, while the insertion loss corresponds to 0.9dB. (b) Insertion loss is plotted as a function of the coupling strength $g_{ac}$ and $g_{bc}$. The other parameters are the same as in Fig. \ref{tu3}(a).}\label{tu7}
\end{figure}

Finally, the insertion loss (IL), defined as $|20*\mathrm{log}_{10}|S_{12}^{M}||$, is another crucial performance index in practical applications. The most desirable performance is high isolation ratio with low insertion loss, which can be realiezd in our system by optimize damping rates $\Gamma_1$ and $\Gamma_2$ for the different optical modes. In Fig.~\ref{tu7}, we plot the transmission spectrum which achieves a performance of about 18dB isolation ratio with 1dB insertion loss, in which the parameters are consistent with those shown in Fig. \ref{tu5}. As shown in Fig.~\ref{tu7}(b), we plot the insertion loss as a function of $\Gamma_1$ and $\Gamma_2$, where the blue dashed line is the contour line of 3dB of insertion loss. Based on this, we can choose suitable parameter ranges to decrease the insertion loss of the system while achieving a larger isolation, which represents the optimal conditions.
\section{Zero damping transmission in Non-Markovian environments}\label{Sec4}
When environmental factors are taken into account, the system dynamics are influenced by interactions with the environment, particularly through dissipation and photon backflow. Dissipation corresponds to the Markovian approximation, while photon backflow reveals non-Markovian effects \cite{Breuer2002,Breuer2104012009,Breuer0210022016,Vega0150012017}. In this section, we investigate the influences of non-Markovian effects on the efficiency of transmission spectrum. First, we derive non-Markovian input-output relationship. For this purpose, we construct a system that the cavity interacts with the non-Markovian environments consisting of a series of bosonic modes (eigenfrequency $x_k$, $n_k$ and $y_k$) \cite{Liu9312011,Xiong0321012019,Cialdi0521042019, Zhang0638532016,Khurana0221072019,Madsen2336012011,Guo2304012021,Li1405012022, Uriri0521072020,Anderson32021993,Liu0622082020,Fanchini2104022014, Haseli0521182014,Goswami0224322021,Debiossac2006012022}. The total Hamiltonian is written as
\begin{align}
\hat{H}=&\hat{H}_{s}+\hat{H}_{I}\label{H_sI-1}
\end{align}
with 
\begin{align}
{{\hat H}_s} =& \ \hbar {\omega _a}{{\hat a}^\dag }\hat a + \hbar {\omega _b}{{\hat b}^\dag }\hat b + \hbar {\omega _c}{{\hat c}^\dag }\hat c \nonumber\\&+ \hbar \left( {{g_{ac}}{{\hat a}^\dag }\hat c + {g_{bc}}{e^{i\phi }}{{\hat b}^\dag }\hat c + J{{\hat a}^\dag }\hat b + {\rm{H}}{\rm{.c}}{\rm{.}}} \right),\label{H_s-1}
\\
{{\hat H}_I} =& \ \hbar \sum\nolimits_k {{x_k}} \hat d_k^\dag {{\hat d}_k} + \hbar \sum\nolimits_k {\left( {{W_k}\hat a\hat d_k^\dag  + W_k^*{{\hat a}^\dag }{{\hat d}_k}} \right)} \nonumber\\
 &+ \hbar \sum\nolimits_k {{n_k}} \hat e_k^\dag {{\hat e}_k} + \hbar\sum\nolimits_k {\left( {{G_k}\hat b\hat e_k^\dag  + G_k^*{{\hat b}^\dag }{{\hat e}_k}} \right)} \nonumber\\
 &+ \hbar\sum\nolimits_k {{y_k}} \hat z_k^\dag {{\hat z}_k} + \hbar\sum\nolimits_k {\left( {{O_k}\hat c\hat z_k^\dag  + O_k^*{{\hat c}^\dag }{{\hat z}_k}} \right)},\label{HI}
\end{align}
where $\hat{a}^\dagger(\hat{a})$, $\hat{b}^\dagger(\hat{b})$ and $\hat{c}^\dagger(\hat{c})$ denote the creation (annihilation) operators of the optical photon modes with frequencies $\omega_a$, $\omega_b$ and $\omega_c$. $J$, $g_{ac}$ and $g_{bc}$ correspond to the coupling strengths. The bosonic modes of the systems interact with the $k$th mode (eigenfrequencies $x_k$, $n_k$ and $y_{k}$) of non-Markovian environments, which are modeled as collections of infinite modes via the creation (annihilation) operators $\hat{d}^\dagger_{k}(\hat{d}_{k})$, $\hat{e}^\dagger_{k}(\hat{e}_{k})$
and $\hat{z}^\dagger_{k}(\hat{z}_{k})$. The coefficients $W_{k}$, $G_k$ and $O_k$ are interacting strengths between three environments and three bosonic modes. With the operators expectation values defined by $a(t) \equiv \langle \hat{a}(t) \rangle$, $b(t) \equiv \langle \hat{b}(t) \rangle$, $c(t) \equiv \langle \hat{c}(t) \rangle$, $d_k(t) \equiv \langle \hat{d}_k(t) \rangle$, $e_k(t) \equiv \langle \hat{e}_k(t) \rangle$ and $z_k(t) \equiv \langle \hat{z}_k(t) \rangle$,
the Heisenberg equations of the systems give
\begin{align}
&\dot a(t) =  - i{\omega _a}a(t) - iJb(t) - i{g_{ac}}c(t) - i\sum\nolimits_k {W_k^*} {d_k}(t),\nonumber\\
&\dot b(t) =  - i{\omega _b}b(t) - iJa(t) - i{g_{bc}}{e^{i\phi }}c(t) - i\sum\nolimits_k {G_k^*} {e_k}(t),\nonumber\\
&\dot c(t) =  - i{\omega _c}c(t) - i{g_{ac}}a(t) - i{g_{bc}}{e^{ - i\phi }}b(t) - i\sum\nolimits_k {O_k^*} {f_k}(t),\nonumber\\
&\dot{d}_k(t)=-ix_{k}d_{k}(t)-iW_{k}a(t), \dot{e}_k(t)=-in_{k}e_{k}(t)-iG_{k}b(t),\nonumber\\
&\dot{z}_k(t)=-iy_{k}z_{k}(t)-iO_{k}c(t). \label{dotabc}
\end{align}
Solving Eq.~\eqref{dotabc}, we obtain solutions of the environments parts for $t\geq0$
\begin{eqnarray}
\begin{aligned}
d_{k}(t)&=d_{k}(0)e^{-ix_{k}t}-iW_{k}\int_{0}^{t}d{\tau}a(\tau)e^{-ix_{k}(t-\tau)},\\
e_{k}(t)&=e_{k}(0)e^{-in_{k}t}-iG_{k}\int_{0}^{t}d{\tau}b(\tau)e^{-in_{k}(t-\tau)},\\
z_{k}(t)&=z_{k}(0)e^{-iy_{k}t}-iO_{k}\int_{0}^{t}d{\tau}c(\tau)e^{-iy_{k}(t-\tau)},
\label{dotdef}
\end{aligned}
\end{eqnarray}
where the first terms on the right-hand side of Eq.~\eqref{dotdef} represent the free-propagating parts of the environments fields, while the
second terms describe the influences of environments on the cavities dynamics. Substituting Eq.~\eqref{dotdef} into Eq.~\eqref{dotabc}, we obtain the integro-differential equations
\begin{align}
\dot{a}(t)  &= -i\omega_{a}a(t)-iJb(t)-ig_{ac}c(t)-K_{1}(t) \nonumber \\
           & \quad -\int_{0}^{t}d{\tau}a(\tau)f_{1}(t-\tau),  \label{dotak}\\
\dot{b}(t)  &= -i\omega_{b}b(t)-iJa(t)-ig_{bc}e^{i\phi}c(t)-K_{2}(t) \nonumber \\
           & \quad -\int_{0}^{t}d{\tau}b(\tau)f_{2}(t-\tau), \label{dotbk}\\
\dot{c}(t)  &= -i\omega_{c}c(t)-ig_{ac}a(t)-ig_{bc}e^{-i\phi}b(t)-K_{3}(t) \nonumber \\
           & \quad -\int_{0}^{t}d{\tau}c(\tau)f_{3}(t-\tau), \label{dotck}
\end{align}
where
\begin{eqnarray}
\begin{aligned}
K_{1}(t)&=i\sum_{k}W_{k}^{*}d_{k}(0)e^{-ix_{k}t}\\
&=\int_{-\infty}^{+\infty}d{\tau}\kappa_{1}^{*}(t-\tau)a_{in}(\tau),
\end{aligned}\\
\begin{aligned}
K_{2}(t)&=i\sum_{k}G_{k}^{*}e_{k}(0)e^{-in_{k}t}\\
&=\int_{-\infty}^{+\infty}d{\tau}\kappa_{2}^{*}(t-\tau)b_{in}(\tau),
\end{aligned}\\
\begin{aligned}
K_{3}(t)&=i\sum_{k}O_{k}^{*}z_{k}(0)e^{-iy_{k}t}\\
&=\int_{-\infty}^{+\infty}d{\tau}\kappa_{3}^{*}(t-\tau)c_{in}(\tau).
\end{aligned}
\end{eqnarray}
The cavities interact with the incoming and outgoing modes of environments at both ends, where we have defined the expectation values of the input field operators
\begin{eqnarray}
a_{in}(t)&=&\frac{-1}{\sqrt{2\pi}}\sum_{k}e^{-ix_{k}t}d_{k}(0),\\
b_{in}(t)&=&\frac{-1}{\sqrt{2\pi}}\sum_{k}e^{-in_{k}t}e_{k}(0),\\
c_{in}(t)&=&\frac{-1}{\sqrt{2\pi}}\sum_{k}e^{-iy_{k}t}f_{k}(0),
\end{eqnarray}
and the impulse response functions
\begin{eqnarray}
\kappa_{1}(t)&=&\frac{i}{\sqrt{2\pi}}\sum_{k}e^{ix_{k}t}W_{k},\\
\kappa_{2}(t)&=&\frac{i}{\sqrt{2\pi}}\sum_{k}e^{in_{k}t}G_{k},\\
\kappa_{3}(t)&=&\frac{i}{\sqrt{2\pi}}\sum_{k}e^{iy_{k}t}O_{k},
\end{eqnarray}
or in the continuum
\begin{eqnarray}
\kappa_{1}(t-{\tau})&=&\frac{i}{\sqrt{2\pi}}\int_{-\infty}^{+\infty}{e^{ix_k{(t-\tau)}}}W(\omega)d\omega,\label{kappaa}\\
\kappa_{2}(t-{\tau})&=&\frac{i}{\sqrt{2\pi}}\int_{-\infty}^{+\infty}{e^{in_k{(t-\tau)}}}G({\omega})d\omega,\label{kappab}\\
\kappa_{3}(t-{\tau})&=&\frac{i}{\sqrt{2\pi}}\int_{-\infty}^{+\infty}{e^{iy_k{(t-\tau)}}}O({\omega})d\omega,\label{kappac}
\end{eqnarray}
where we have made the replacements by $W_k\rightarrow{W(\omega)}$, $G_k\rightarrow{G(\omega)}$, $O_k\rightarrow{O(\omega)}$. The correlation function $f_1(t)$, $f_2(t)$, and $f_3(t)$ in Eqs.~(\ref{dotak})-(\ref{dotck}) are given by
\begin{eqnarray}
\begin{aligned}
f_{v}(t)=&\int_{-\infty}^{+\infty}{J_{v}(\omega)e^{-i\omega{t}}}d\omega\\
=&\int_{-\infty}^{+\infty}\kappa_v(-\zeta)\kappa^*_v(t-\zeta)d\zeta,(v=1,2,3)\label{fa}
\end{aligned}
\end{eqnarray}
where
$J_{1}(\omega)=|W_{k}|^{2},$
$J_{2}(\omega)=|G_{k}|^{2},$
and $J_{3}(\omega)=|O_{k}|^{2}$ represent the spectral densities of the different environments. $f_1(t)$, $f_2(t)$, and $f_3(t)$ describe the non-Markovian fluctuation-dissipation relationship of the environments. Similarly, we obtain solutions of the environments parts for $t_1\geq t$
\begin{eqnarray}
\!\!\!\!d_{k}(t) \!=\! d_{k}(t_1)e^{-ix_{k}(t-t_{1})} + iW_{k}\int_{t}^{t_1}d{\tau}a(\tau)e^{-ix_{k}(t-\tau)},\\
\!\!\!\!\!\!\!e_{k}(t)\!=\!e_{k}(t_1)e^{-in_{k}(t-t_{1})} + iG_{k}\int_{t}^{t_1}d{\tau}b(\tau)e^{-in_{k}(t-\tau)},\\
\!\!\!\!\!\!\!z_{k}(t)\!=\!z_{k}(t_1)e^{-iy_{k}(t-t_{1})} + iO_{k}\int_{t}^{t_1}d{\tau}c(\tau)e^{-iy_{k}(t-\tau)},
\end{eqnarray}
and the integro-differential equations
\begin{align}
\dot{a}(t)  &= -i\omega_{a}a(t)-iJb(t)-ig_{ac}c(t)-K_{1}'(t) \nonumber \\
           & \quad +\int_{t}^{t_1}d{\tau}a(\tau)f_{1}(t-\tau),  \label{dotaK}\\
\dot{b}(t)  &= -i\omega_{b}b(t)-iJa(t)-ig_{bc}e^{i\phi}c(t)-K_{2}'(t) \nonumber \\
           & \quad +\int_{t}^{t_1}d{\tau}b(\tau)f_{2}(t-\tau), \label{dotbK}\\
\dot{c}(t)  &= -i\omega_{c}c(t)-ig_{ac}a(t)-ig_{bc}e^{-i\phi}b(t)-K_{3}'(t) \nonumber \\
           & \quad +\int_{t}^{t_1}d{\tau}c(\tau)f_{3}(t-\tau), \label{dotcK}
\end{align}
where
\begin{eqnarray}
\begin{aligned}
K_{1}'(t)&=i\sum_{k}W_{k}^{*}d_{k}(t_{1})e^{-ix_{k}(t-t_{1})}\\
&=\int_{-\infty}^{+\infty}d{\tau}\kappa_{1}^{*}(t-\tau)a_{out}(\tau),
\end{aligned}\\
\begin{aligned}
K_{2}'(t)&=i\sum_{k}G_{k}^{*}e_{k}(t_{1})e^{-in_{k}(t-t_{1})}\\
&=\int_{-\infty}^{+\infty}d{\tau}\kappa_{2}^{*}(t-\tau)b_{out}(\tau),
\end{aligned}\\
\begin{aligned}
K_{3}'(t)&=i\sum_{k}O_{k}^{*}z_{k}(t_{1})e^{-iy_{k}(t-t_{1})}\\
&=\int_{-\infty}^{+\infty}d{\tau}\kappa_{3}^{*}(t-\tau)c_{out}(\tau),
\end{aligned}
\end{eqnarray}
with
\begin{eqnarray}
a_{out}(t)&=&\frac{-1}{\sqrt{2\pi}}\sum_{k}e^{-ix_{k}(t-t_{1})}d_{k}(t_1),\\
b_{out}(t)&=&\frac{-1}{\sqrt{2\pi}}\sum_{k}e^{-in_{k}(t-t_{1})}e_{k}(t_1),\\
c_{out}(t)&=&\frac{-1}{\sqrt{2\pi}}\sum_{k}e^{-iy_{k}(t-t_{1})}f_{k}(t_1).
\end{eqnarray}
By comparing Eqs.~\eqref{dotak}-\eqref{dotck} with Eqs.~\eqref{dotaK}-\eqref{dotcK}, the non-Markovian input-output relations can be expressed as (taking $t_1\rightarrow{t}$)
\begin{eqnarray}
\begin{aligned}
a_{out}(t)-a_{in}(t)=&\int_{0}^{t}d\tau\kappa_{1}(\tau-t)a(\tau),\\
b_{out}(t)-b_{in}(t)=&\int_{0}^{t}d\tau\kappa_{2}(\tau-t)b(\tau),\\
c_{out}(t)-c_{in}(t)=&\int_{0}^{t}d\tau\kappa_{3}(\tau-t)c(\tau),\label{shut}
\end{aligned}
\end{eqnarray}
where $\kappa_v(t-\tau)(v=1,2,3)$ is defined by Eqs.~\eqref{kappaa}-\eqref{kappac}. Here, impulse response functions can be set as
$\kappa_\nu(t)=i\lambda_\nu\sqrt{\Gamma_{\nu}}e^{\lambda_\nu t}\theta(-t)$
($\theta(t)$ is the unit step function, where $\theta(t)$ = 1 for $t\geq{0}$ otherwise $\theta(t)=0$), which leads to correlation function $f_\nu(t)=\frac{1}{2}\lambda_\nu\Gamma_{\nu}e^{-\lambda_\nu|t|}(v=1,2,3)$. According to Eq.~({\ref{kappaa}})-({\ref{kappac}}), the spectral response functions are $W(\omega)=\frac{1}{\sqrt{2\pi}}\int_{-\infty}^{
+\infty}e^{-i\omega t}\kappa_1(t)dt$, $G(\omega)=\frac{1}{\sqrt{2\pi}}\int_{-\infty}^{
+\infty}e^{-i\omega t}\kappa_2(t)dt$ and $O(\omega)=\frac{1}{\sqrt{2\pi}}\int_{-\infty}^{
+\infty}e^{-i\omega t}\kappa_3(t)dt$. Applying the Fourier transform to Eqs.~\eqref{kappaa}-\eqref{kappac}, we obtain
\begin{eqnarray}
W(\omega)&=&\sqrt{\frac{\Gamma_{1}}{2\pi}}\frac{\lambda_{1}}{\lambda_{1}-i\omega},\label{ga}\\
G(\omega)&=&\sqrt{\frac{\Gamma_{2}}{2\pi}}\frac{\lambda_{2}}{\lambda_{2}-i\omega},\label{gb}\\
O(\omega)&=&\sqrt{\frac{\Gamma_{3}}{2\pi}}\frac{\lambda_{3}}{\lambda_{3}-i\omega},\label{gc}
\end{eqnarray}
where $\lambda_{v}(v = 1, 2, 3)$ is the environmental spectrum width, while $\Gamma_{v}(v = 1, 2, 3)$ is defined as the disspation of the environmental modes acting on the cavities. Thus the spectral density of the environments reads
\begin{eqnarray}
J_{v}(\omega)=\frac{\Gamma_{v}}{2\pi}\frac{\lambda_{v}^{2}}{\lambda_{v}^{2}+\omega^{2}},(v=1,2,3),
\label{J}
\end{eqnarray}
which corresponds to a Gaussian Ornstein-Uhlenbeck process~\cite{Uhlenbeck8231930,Gillespie20841996,Jing2404032010}. Through pseudomode theory \cite{Barnett1997,Garraway22901997,Garraway46361997,Man0621042014,Man57632015,Mazzola0121042009,
Pleasance0621052017}, we can give the concrete form of impulse response functions $\kappa_\nu(t)$ and correlation function $f_\nu(t)$ (see Appendix~\ref{APP.1}). Specially, in the wideband limit, which means $\lambda_{v}$ tends to infinity (i.e., $\lambda_{v} \rightarrow{\infty}$), the environments become memoryless. The spectral density approximately
takes $J_{v}(\omega)\rightarrow\sqrt{\Gamma_{v}/{2\pi}}$, or equivalently, $W(\omega)\rightarrow\sqrt{\Gamma_1/{2\pi}}$, $G(\omega)\rightarrow\sqrt{\Gamma_2/{2\pi}}$
, $O(\omega)\rightarrow\sqrt{\Gamma_3/{2\pi}}$. According to these, we have $f_{v}(t) = \Gamma_{v}\delta(t)$ and $\kappa_{v}(t)=i\sqrt{\Gamma_{v}}\delta(t)$. Substituting the results into Eq.~(\ref{shut}), we can obtain the Markovian input-output relations
\begin{eqnarray}
a_{out}(t)-a_{in}(t)&=&i\sqrt{\Gamma_1}a(t),\label{aout}\\
b_{out}(t)-b_{in}(t)&=&i\sqrt{\Gamma_2}b(t),\label{bout}\\
c_{out}(t)-c_{in}(t)&=&i\sqrt{\Gamma_3}c(t).\label{cout}
\end{eqnarray}
Here, we demonstrate that the Markovian input-output relations, as described by Eqs.~\eqref{aout}-\eqref{cout}, are equivalent to those defined in Refs.~\cite{Gardiner2000,Walls1994,Scully1997} and can return to the results of Refs.~\cite{Gardiner2000,Walls1994,Scully1997} by the replacements $W_{k}\rightarrow{iW(\omega)}$, $G_{k}\rightarrow{iG(\omega)}$,
$O_{k}\rightarrow{iO(\omega)}$ in Eqs.~\eqref{HI} and ~\eqref{kappaa}-\eqref{kappac}, the impulse response functions in the continuum are rewritten as
\begin{eqnarray}
\kappa_{1}(t-{\tau})&=&-\frac{1}{\sqrt{2\pi}}\int_{-\infty}^{+\infty}{e^{ix_k{(t-\tau)}}}W(\omega)d\omega,\label{kappaai}\\
\kappa_{2}(t-{\tau})&=&-\frac{1}{\sqrt{2\pi}}\int_{-\infty}^{+\infty}{e^{in_k{(t-\tau)}}}G({\omega})d\omega,\label{kappabi}\\
\kappa_{3}(t-{\tau})&=&-\frac{1}{\sqrt{2\pi}}\int_{-\infty}^{+\infty}{e^{iy_k{(t-\tau)}}}O({\omega})d\omega.\label{kappaci}
\end{eqnarray}
Substituting the Eqs.~\eqref{ga}-\eqref{gc} into Eqs.~\eqref{kappaai}-\eqref{kappaci}, we obtain
\begin{eqnarray}
\kappa_{v}(\tau-t)&=&-\sqrt{\Gamma_{v}}\lambda_{v}e^{-\lambda_{v}(t-\tau)}\theta(t-\tau),
\end{eqnarray}
so we get $\kappa_{v}(t)=-\sqrt{\Gamma_{v}}\delta(t)$ in the wideband limit (i.e., $\lambda_{v} \rightarrow{\infty}$). According to Eq.~({\ref{shut}}), we obtain the Markovian input-output relations, which return to Eqs.~(\ref{Main}) and ~(\ref{Mbin}).

\subsection{Adiabatic elimination with non-Markovian effects}
Adiabatic elimination is a common method for dealing with multi-degree-of-freedom problems. It can help us simplify complex systems to understand their underlying physics, and provide a theoretical tool for controlling quantum systems. However, when non-Markovian effects are considered, the system's dynamics become complex due to interactions with the environment, and the adiabatic elimination has not been explored so far. Therefore, we extend the adiabatic elimination method to the non-Markovian regime. According to Eq.~(\ref{dotck}), we can get
\begin{align}
&\dot{C}=-ig_{ac}A(t)e^{-i\Delta_{ac}t}-ig_{bc}e^{-i\phi}B(t)e^{-i\Delta_{bc}t}\label{CJ}\\
&-i\sum_{k}e^{-i(y_{k}-\omega_{c})t}O_{k}^{*}z_{k}(0)\!\!-\!\!\int_{0}^{t}c(\tau)e^{-i\omega_{c}(\tau-t)}f_{3}(t-\tau)d\tau.\nonumber
\end{align}
Since Eq.~(\ref{CJ}) contains an integral term, we can expand it into a series and ignore higher-order terms to obtain an approximate solution
\begin{eqnarray}
\begin{aligned}
&\int_{0}^{t}C(\tau)e^{-i\omega_{c}(\tau-t)}\frac{1}{2}\Gamma_{3}\lambda_{3}e^{-\lambda_3(t-\tau)}d\tau\\
&=~C(\tau)\frac{1}{2}\Gamma_{3}\lambda_{3}\frac{1}{\lambda_{3}-i\omega_c}e^{-\lambda_3(t-\tau)-i\omega_{c}(\tau-t)}\\
&-\int_0^{t}(C(\tau)\frac{1}{2}\Gamma_{3}\lambda_{3}\frac{1}{\lambda_{3}-i\omega_c})'e^{-\lambda_3(t-\tau)-i\omega_{c}(\tau-t)}d\tau.\label{JF}
\end{aligned}
\end{eqnarray}
For Eq.~(\ref{JF}), we only take the first-order terms and get
\begin{eqnarray}
\begin{aligned}
&\int_{0}^{t}C(\tau)e^{-i\omega_{c}(\tau-t)}\frac{1}{2}\Gamma_{3}\lambda_{1}e^{-\lambda(t-\tau)}d\tau\\
&\approx
C(t)\frac{1}{2}\Gamma_{3}\lambda_{3}\frac{1}{\lambda_{3}-i\omega_c},
\end{aligned}
\end{eqnarray}
which leads to Eq.~(\ref{CJ}) becoming
\begin{align}
&\dot{C}(t)=-ig_{ac}A(t)e^{-i\Delta_{ac}t}-ig_{bc}e^{-i\phi}B(t)e^{-i\Delta_{bc}t}\nonumber\\
&-i\sum_{k}e^{-i(y_{k}-\omega_{c})t}O_{k}^{*}z_{k}(0)-C(t)\frac{1}{2}\Gamma_{3}\lambda_{3}\frac{1}{\lambda_{3}-i\omega_c}.\label{nondotc}
\end{align}
On account of $z_{k}(0)=0$, the solution to Eq.~(\ref{nondotc}) can be written as
\begin{eqnarray}
\begin{aligned}
C(t)=&-ig_{ac}\int_0^{t}dt'A(t')e^{-i\Delta_{ac}t'}e^{-h(t-t')}\\
&-ig_{bc}e^{-i\phi}\int_0^{t}dt'B(t')e^{-i\Delta_{bc}t'}e^{-h(t-t')},
\label{C3JF}
\end{aligned}
\end{eqnarray}
or
\begin{eqnarray}
\begin{aligned}
C(t)=&-\frac{ig_{ac}}{h-i\Delta_{ac}}A(t)e^{-i\Delta_{ac}t}\\
&-\frac{ig_{bc}e^{-i\phi}}{h-i\Delta_{bc}}B(t)e^{-i\Delta_{bc}t},
\label{C3F}
\end{aligned}
\end{eqnarray}
where $h=\frac{1}{2}\Gamma_{3}\lambda_{3}\frac{1}{\lambda_{3}-i\omega_c}$. Combining Eqs.~(\ref{dotak}) and~(\ref{dotbk}), the adiabatic elimination Heisenberg equations with non-Markovian effects are rewritten as
\begin{align}
\dot{a}(t)=&-i\omega_{a}a(t)-iJb(t)-F_2a(t)-F_4b(t)-K_{1}(t)\nonumber\\
&-\int_{0}^{t}d{\tau}a(\tau)f_{1}(t-\tau),\label{dotakJ}\\
\dot{b}(t)=&-i\omega_{b}b(t)-iJa(t)-F_1b(t)-F_3a(t)-K_{2}(t)\nonumber\\
&-\int_{0}^{t}d{\tau}b(\tau)f_{2}(t-\tau),\label{dotbkJ}
\end{align}
where 
\begin{align}
F_1=&\frac{g_{bc}^{2}}{h-i\Delta_{bc}},
\ F_2=\frac{g_{ac}^{2}}{h-i\Delta_{ac}},\nonumber\\
F_3=&\frac{g_{ac}g_{bc}e^{i\phi}}{h-i\Delta_{ac}},
\ F_4=\frac{g_{ac}g_{bc}e^{-i\phi}}{h-i\Delta_{bc}}\label{F1-F4}
\end{align}
Making a modified Laplace transformation \cite{Uchiyama021128, Saeki031131,Shen052122} $\zeta(\omega)=\int_{0}^{\infty}e^{i\omega{t}}\zeta(t)dt$, where $e^{i\omega{t}}\rightarrow{e^{i\omega{t}-\epsilon{t}}}$ with
$\epsilon\rightarrow{0^+}$ make $\zeta(\omega)$ converge to a finite value, the cavities equations satisfy
\begin{align}
-i\omega{a}(\omega)&=-i\omega_{a}a(\omega)-iJb(\omega)-F_2a(\omega)-F_4b(\omega)\nonumber\\
&-\widetilde{\kappa}_{1}(\omega)[a_{in}(\omega)-a_{in}(i\lambda_{1})]-a(\omega)f_{1}(\omega),\label{oa}\\
-i\omega{b}(\omega)&=-i\omega_{b}b(\omega)-iJa(\omega)-F_1b(\omega)-F_3a(\omega)\nonumber\\
&-\widetilde{\kappa}_{2}(\omega)[b_{in}(\omega)-b_{in}(i\lambda_{2})]-b(\omega)f_{2}(\omega),\label{ob}
\end{align}
with
\begin{align}
\!\!\!\widetilde{\kappa}_{v}(\omega)\!=\!\int_{-\infty}^{0}\kappa_{v}^{*}(t')e^{i\omega{t'}}dt', f_{v}(\omega)\!=\!\int_{0}^{\infty}f_{v}(t')e^{i\omega{t'}}dt'.\!\!\!
\end{align}
According to Eqs.~({\ref{oa}}) and~({\ref{ob}}), we analyze the impact of $a_{in}(i\lambda_1)$ and $b_{in}(i\lambda_2)$ by defining the ratio $\Theta_1(\lambda_1,\omega)=a_{in}(i\lambda_1)/a_{in}(\omega)$ and $\Theta_2(\lambda_2,\omega)=b_{in}(i\lambda_2)/b_{in}(\omega)$, which are inhomogeneous terms depending on the specific forms of the input field $a_{in}(i\lambda_1)$ and $b_{in}(i\lambda_2)$. We set the input field as damped oscillation $a_{in}(t)=ve^{-{\eta}t}$sin$(ut^2)$ for $\eta>0$ and $u>0$, and Gaussian profile $a_{in}(t)=ve^{-{\eta}t^2}$cos$(ut)$ for $\eta>0$ and $u>0$, which respectively correspond to
\begin{widetext}
\begin{align}
{\Theta _1}({\lambda _1},\omega ) &= \frac{{\cos \left[ {\frac{{{{({\lambda _1} + \eta )}^2}}}{{4u}}} \right]\left[ {1 - 2hc(\frac{{{\lambda _1} + \eta }}{{\sqrt {2\pi u} }})} \right] + \sin \left[ {\frac{{{{({\lambda _1} + \eta )}^2}}}{{4u}}} \right]\left[ {1 - 2hs(\frac{{{\lambda _1} + \eta }}{{\sqrt {2\pi u} }})} \right]}}{{\cos \left[ {\frac{{{{(\eta  - i\omega )}^2}}}{{4u}}} \right]\left[ {1 - 2hc(\frac{{\eta  - i\omega }}{{\sqrt {2\pi u} }})} \right] + \sin \left[ {\frac{{{{(\eta  - i\omega )}^2}}}{{4u}}} \right]\left[ {1 - 2hs(\frac{{\eta  - i\omega }}{{\sqrt {2\pi u} }})} \right]}},\label{erhi-1}\\
{\Theta _1}({\lambda _1},\omega ) &= \exp \left[ {\frac{{\lambda _1^2 + {\omega ^2} + 2u\left( {\omega  - i{\lambda _1}} \right)}}{{4\eta }}} \right]\left\{ {\frac{{i + {e^{\frac{{iu{\lambda _1}}}{\eta }}}\left[ {i + erhi\left( {\frac{{u - i{\lambda _1}}}{{2\sqrt \eta  }}} \right)} \right] - erhi\left( {\frac{{u + i{\lambda _1}}}{{2\sqrt \eta  }}} \right)}}{{i + {e^{\frac{{u\omega }}{\eta }}}\left[ {i + erhi\left( {\frac{{u - \omega }}{{2\sqrt \eta  }}} \right)} \right] - erhi\left( {\frac{{u + \omega }}{{2\sqrt \eta  }}} \right)}}} \right\},\label{erhi-2}
\end{align}
\end{widetext}
where $hc(l)=\int_{0}^{l}$cos$(\pi{t^2}/2)dt$, $hs(l)=\int_{0}^{l}$sin$(\pi{t^2}/2)dt$, $erhi(l)=erhi(il)/i$ with $erh(l)=\frac{2}{\sqrt{\pi}}\int_{0}^{l}e^{-t^2}dt$.
For input fields with damped oscillation and Gaussian profiles, we find that $\Theta_1(\lambda_1,\omega)$ arises from non-Markovian effects and lacks Markovian correspondent. Specifically, for the concrete input field and given parameters, e.g., $\eta=0.0015\omega_a$, $u =0.004\omega_a$, $|\Theta_1(\lambda_1,\omega)|\sim 10^{-5}$ for $\lambda_1=\omega_a$
which is in non-Markovian regime and $|\Theta_1(\lambda_1,\omega)|\sim 10^{-8}$ for $\lambda_1=11\omega_a$ which means weak non-Markovian effects at the interval $\omega\in(0.9\omega_a,1.2\omega_a)$. Under the Markovian approximation as $\lambda_1\rightarrow\infty$, $\Theta_1(\lambda_1,\omega)$ tends to zero, where the input-field dependent term $a_{in}(i\lambda_1)$ is shown to be negligible compared to the other terms. Consequently, the influences of the inhomogeneous terms on system operators is not taken into account in the following studies of nonreciprocal transmission. For $b_{in}(i\lambda_2)$, we can also demonstrate this by using the same approach.

Considering the above , we consider a two-side cavity and the Heisenberg-Langvan equations of the system are
\begin{eqnarray}
\begin{aligned}
-i\omega{a}(\omega)=&-i\omega_{a}a(\omega)-iJb(\omega)-F_2a(\omega)-F_4b(\omega)\\
&-\widetilde{\kappa}_{1}(\omega)p_{in}(\omega)-a(\omega)f_{1}(\omega),
\end{aligned}\\
\begin{aligned}
-i\omega{b}(\omega)=&-i\omega_{b}a(\omega)-iJa(\omega)-F_1b(\omega)-F_3a(\omega)\\
&-\widetilde{\kappa}_{2}(\omega)p_{in}(\omega)-b(\omega)f_{2}(\omega),
\end{aligned}
\end{eqnarray}
the non-Markovian input-output relations transmitted from left to right in the frequency regime is expressed as
\begin{eqnarray}
p_{out}(\omega)-p_{in}(\omega)\approx a(\omega)\kappa_{1}(-\omega).
\end{eqnarray}
Similarly, we can obtain the non-Markovian input-output relations transmitted from right to left in the frequency regime
\begin{eqnarray}
q_{out}(\omega)-q_{in}(\omega)\approx a(\omega)\kappa_{1}(-\omega).
\end{eqnarray}
The input and output fields are related through
\begin{eqnarray}
\begin{aligned}
\mathbf{M_{out}(\omega)}=\mathbf{S^{NM}(\omega)}\mathbf{M_{in}(\omega)},
\end{aligned}
\end{eqnarray}
or
\begin{eqnarray}
\begin{matrix}
\mathbf{S^{NM}(\omega)}=\left[
\begin{array}{cc}
 S_{12}^{NM} & S_{11}^{NM} \\
 S_{22}^{NM} & S_{21}^{NM} \\
\end{array}
\right],
\end{matrix}
\end{eqnarray}
with $\mathbf{M_ {out}(\omega)}=[p_{out}(\omega),q_{out}(\omega)]^T$ and $\mathbf{M_{in}(\omega)}=[p_{in}(\omega),q_{in}(\omega)]^T$. $\mathbf{S^{NM}(\omega)}$ denotes the scattering matrix of the system and $S_{12,(21)}^{NM}$ describes the photon transmission of the signal from left (right) to right (left) in the system., so we calculate the transmission coefficient with $\phi$ taking the value of $0$ for $S_{12}^{NM}$ and $\pi$ for $S_{21}^{NM}$
\begin{widetext}
\begin{eqnarray}
\begin{aligned}
S_{12|\phi=0}^{NM}&=1-\frac{-\widetilde{\kappa}_{1}(\omega)[i\omega-i\omega_b-\frac{g_{bc}^2}{h-i\Delta_{bc}}-f_2(\omega)]{\kappa}_{1}(-\omega)}
{[i\omega-i\omega_a-\frac{g_{ac}^2}{h-i\Delta_{ac}}-f_1(\omega)][i\omega-i\omega_b-\frac{g_{bc}^2}{h-i\Delta_{bc}}-f_2(\omega)]+(J-\frac{ig_{ac}g_{bc}e^{i\phi}}{h-i\Delta_{ac}})(J-\frac{ig_{ac}g_{bc}e^{-i\phi}}{h-i\Delta_{bc}})},\\
S_{21|\phi=\pi}^{NM}&=1-\frac{-\widetilde{\kappa}_{1}(\omega)[i\omega-i\omega_b-\frac{g_{bc}^2}{h-i\Delta_{bc}}-f_2(\omega)]{\kappa}_{1}(-\omega)}
{[i\omega-i\omega_a-\frac{g_{ac}^2}{h-i\Delta_{ac}}-f_1(\omega)][i\omega-i\omega_b-\frac{g_{bc}^2}{h-i\Delta_{bc}}-f_2(\omega)]+(J-\frac{ig_{ac}g_{bc}e^{i\phi}}{h-i\Delta_{ac}})(J-\frac{ig_{ac}g_{bc}e^{-i\phi}}{h-i\Delta_{bc}})}.\label{NonF}
\end{aligned}
\end{eqnarray}
\end{widetext}
\section{DISCUSSION OF TRANSMISSION CHARACTERISTICS IN non-Markovian regime}\label{Sec5}
\begin{figure}[t]
\centerline{
\includegraphics[width=8.4cm, height=5.8cm, clip]{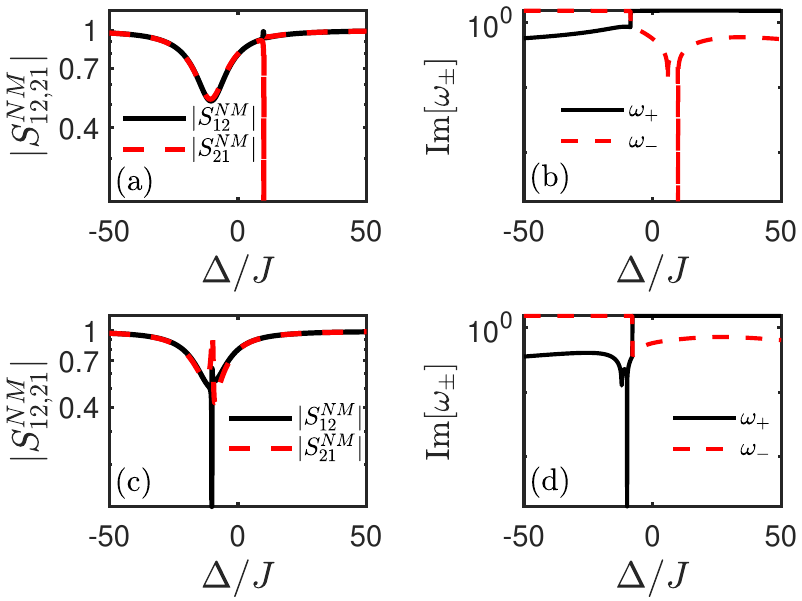}}
\vspace{-0.3cm}
\caption{Transmission spectrum $|S_{21,12}^{NM}|$ in Eq.~(\ref{NonF}) is plotted as a function of $\Delta$ for different direction. (a) Transmission spectrum of the system with $\omega_b-\omega_a=10J$. (b) The logarithm of the imaginary part of the system eigenvalues with $\phi=\pi$. (c) Transmission spectrum of the system with $\omega_b-\omega_a=-10J$. (d) The logarithm of the imaginary part of the system eigenvalues with $\phi=0$. The red dashed line and black solid line represent different eigenvalues, respectively. The parameters chosen are $\omega_a=500J$, $\omega_c=500J$, $\Gamma_1=10J$, $\Gamma_2=0.1J$, $\Gamma_3=100J$, $g_{ac}=20J$, $g_{bc}=2J$, $\lambda_1=\omega_a$, $\lambda_2=\omega_b$, and $\lambda_3=\omega_c$.}\label{tu8}
\end{figure}
In Fig.~\ref{tu8}(a,c), we show transmission spectrum of the system for given spectral widths of the environments ($\lambda_1 = \omega_a$, $\lambda_2 = \omega_b$, $\lambda_3 = \omega_c$). The black line represents the input of the light field from the left side, while the red line denotes the input from the right side. In Fig.~\ref{tu8}(b,d), we plot the logarithmic figure of the imaginary part of the eigenvalues for the effective Hamiltonian as a function of $\Delta$. We can find an ultra sharp dip at $\Delta=10J$ or $\Delta=-10J$ corresponding to the minimum, which means the imaginary part of one eigenvalue is zero. Meanwhile, it shows the nonreciprocity of transmission at the same frequency position corresponding to nonreciprocal transmission with zero reflection in Fig.~\ref{tu8}(a,c). That is to say, there are still nonreciprocal phenomena with the non-Markovian effects, however, the maximum transmission efficiency has changed. Compared to the Markovian cases, transmission efficiency varies in non-Markovian environments due to the dissipation or the backflow oscillation of the photons from the non-Markovian environments. These results are calculated by using the parameters $g_{ac}= 20J$ and $g_{bc}= 2J$.

\begin{figure}[t]
\centerline{
\includegraphics[width=8.2cm, height=5.8cm, clip]{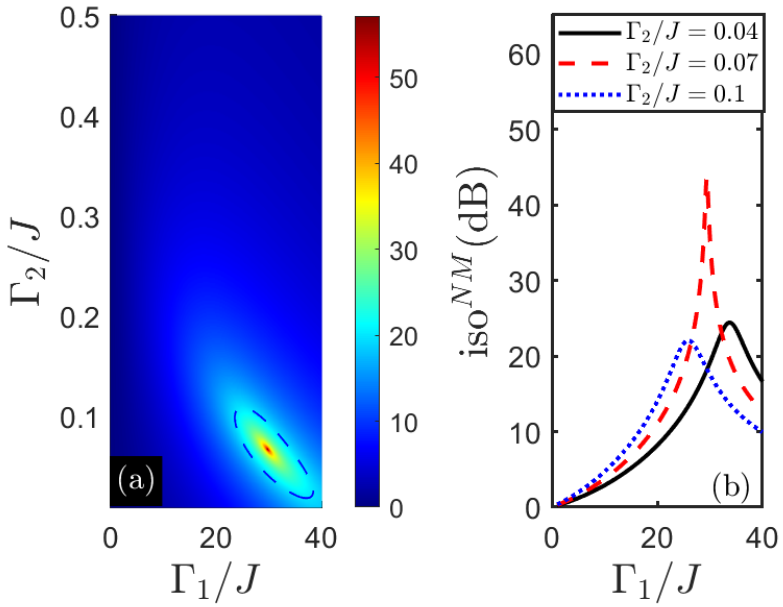}}
\vspace{-0.3cm}
\caption{The influences of external decay on isolation ratio. (a) Isolation ratio according to Eq.~(\ref{NonF}) is plotted as a function of the decay $\Gamma_1$ and $\Gamma_2$ in non-Markovian regime. (b) The isolation ratio is plotted with different $\Gamma_2$ by using Eq.~(\ref{NonF}). The parameters chosen are $\omega_b=510J$, $\Delta=10J$ and the other parameters chosen are the same as in Fig. \ref{tu8}.}\label{tu9}
\end{figure}

For the situation discussed in Fig.~\ref{tu9}, the isolation ratio as a function of the coupling strength $\Gamma_1$ and $\Gamma_2$ is plotted in Fig.~\ref{tu9} with non-Markovian environments, which is calculated by fixing the decay $g_{ac}= 20J$ and $g_{bc}= 2J$. The black dotted line is the contour lines of 20dB of isolation ratio, while the isolation ratio above 50dB can be achieved. Comparing that to the Markovian regime, the isolation ratio reaches the maximum value near $\Gamma_1= 30J$ and $\Gamma_2= 0.07J$. In this case, we also find that the maximum value of the isolation ratio decreases, but the maximum of the tunable range increases. That is to say, the selection for high isolation ratio exhibits a wider range compared to the Markovian case. As shown in Fig.~\ref{tu9}(b), the isolation ratio is more sensitive to change in $\Gamma_2$ than $\Gamma_1$ and it would vary dramatically with the difference of $0.1J$ in $\Gamma_2$, as illustrated in red dotted line. This is consistent with the Markovian regime.

In Fig.~\ref{tu10}(a), with the coupling strengths $g_{ac}=20J$ and $g_{bc}=2J$, we show transmission efficiency as a function of different spectral widths $\lambda_1$ and $\lambda_2$ of the environments. We find that the regime of maximum isolation ratio appears in the lower right corner of the figure with the non-Markovian effects. Interestingly, the maximum value does not occur when the system returns to the Markovian regime (i.e., $\lambda_1 = 20\omega_a$ and $\lambda_2 = 20\omega_b$) in which this would be explained below. In other words, within the lower right corner, the system exhibits higher isolation ratio in the non-Markovian environments. With the purpose of seeing the influences of the environmental spectrum width on transmission spectrum more clearly, the isolation ratio as a function of both $\lambda_1$ and $\lambda_2$ is shown in Fig.~\ref{tu10}(b). As the environmental spectrum widths $\lambda_1$ and $\lambda_2$ further increase, the isolation ratio reaches its maximum value corresponding to the environmental spectrum widths $\lambda_1=4\omega_a$ and $\lambda_2 = 2\omega_b$ as shown by the red dotted line.
\begin{figure}[t]
\centerline{
\includegraphics[width=8.2cm, height=5.8cm, clip]{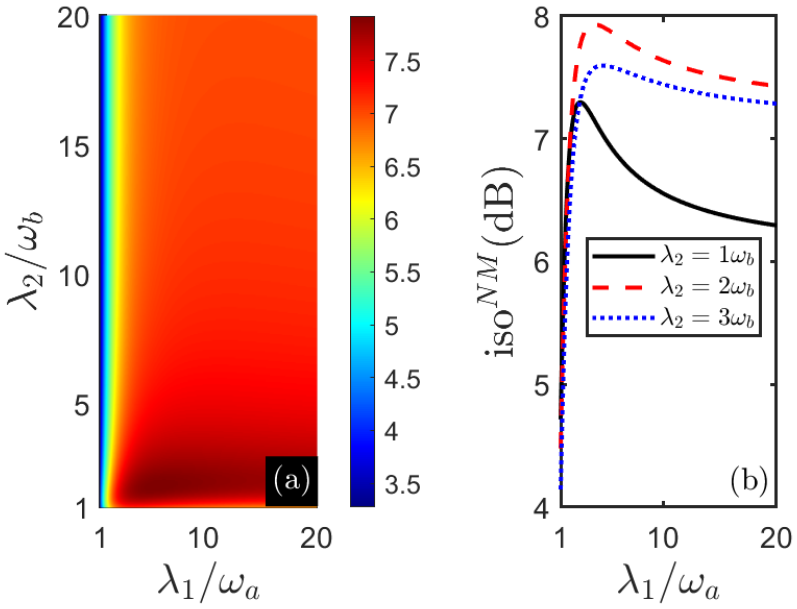}}
\vspace{-0.3cm}
\caption{Analyzing the influences of spectral widths of the environments on nonreciprocity. (a) Isolation ratio by using Eq.~(\ref{NonF}) is plotted as a function of spectral widths $\lambda_1$ and $\lambda_2$ of the environments in non-Markovian regimes. (b) The isolation ratio is plotted with different $\lambda_2$ according to Eq.~(\ref{NonF}). The parameter chosen are $\lambda_3=20\omega_c$, $\omega_b=510J$, $\Delta=10J$ and other parameters are the same as in Fig. \ref{tu8}.}\label{tu10}
\end{figure}

\begin{figure}[t]
\centerline{
\includegraphics[width=0.46\textwidth]{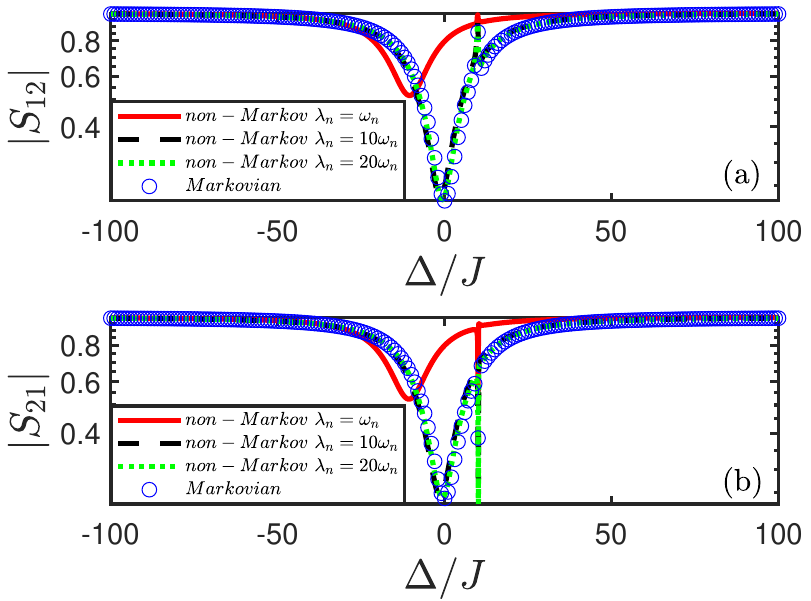}}
\vspace{-0.3cm}
\caption{Comparisons of transmission spectrum calculated by Eqs.~(\ref{S2}) and~(\ref{NonF}) with and without Markovian approximations. Transmission spectrum as a function of $\Delta$ for different direction with different spectral widths of the environments. The figure is plotted with the non-Markovian effects, while the circles indicate the case of Markovian approximations. The parameters chosen are $\Delta=10J$, $\omega_b=510J$ and the other parameters are the same as in Fig.~\ref{tu8}.}\label{tu11}
\end{figure}In order to verify the validity of our method, we simulate our system under the large environmental
spectrum width scenario and find that it is consistent with the Markovian environments.
As shown in Fig.~\ref{tu11}, we plot the transmission of the system with different spectral widths,
where other parameters are chosen as in Fig.~\ref{tu3}. We can see that when the environmental
spectrum width is $\omega_n(n=a,b,c)$, significant differences exist between the non-Markovian
and Markovian cases, as shown by the red solid line. As the
spectrum widths of the environments increase, the efficiency of transmission increases. For the sake of clarity, we separately plot the non-Markovian limit and the Markovian approximation cases, where the environmental spectrum widths taking $\lambda_1=20\omega_a$, $\lambda_2=20\omega_b$, and $\lambda_3=20\omega_c$ are enough for our scheme. We can see the non-Markovian transmission spectrum is consistent with the Markovian transmission spectrum, which means that the systems go back to Markovian regime again.

At the end of the article, let's evaluate the implementation of our scheme in superconducting circuit systems. Specifically, the main part of the circuit consists of three optical crystal nanobeam cavities, of which transmission in different directions can be excited by two separate optical coupling paths and the phase in different directions can be controlled by the pump field with phase shifters. In view of these above discussions, we propose that experimental implementation is feasible.

\section{conclusion}\label{Sec6}
In summary, we have demonstrated three coupled optical modes scheme for achieving nonreciprocal transmission in this paper, which is accomplished by designing a system that exploits the synergistic effects of coherent and dissipative coupling. The scheme gets zero-damping transmission and exhibits unidirectional invisibility. The model explicitly considers the direction-dependent relative phase between coherent and dissipative coupling, breaking time-reversal symmetry of propagation. Moreover, we extend the adiabatic elimination method into the non-Markovian regime and derive the effective Hamiltonian with the non-Markovian effects through the non-Markovian adiabatic elimination method. As a result of analysis, we have shown that nonreciprocal transport persists in the non-Markovian regime driven by the interference of coherent and dissipative couplings. Our scheme provides flexible controllability for non-reciprocal dynamics, which is easily tuned via parameters to optimize performance for high nonreciprocity and low insertion loss.

The zero-reflection nonreciprocal transmission implemented in this paper can protect quantum devices from backscattering, which is very important for the construction of quantum networks. At the same time, our discussion of non-Markovian effects provides a reference for the behavior of nonreciprocity under the influences of environments. We will more explicitly unveil the physical mechanisms governing the different system¡¯s behavior in the non-Markovian environments by applying the studied non-Markovian adiabatic elimination method. As an outlook, we can also explore the fundamental physical system with the Rabi model \cite{Chen033603,Chen043711, Chen52024}, the non-rotating-wave interaction Hamiltonian between the optical cavity and environments $\sum_kJ(\hat{m}+\hat{m}^{\dagger})(\hat{x}_k+{\hat{x}}^{\dagger}_k)$ \cite{Shen033805}, and
anisotropic non-rotating-wave interaction Hamiltonian $\sum_{k}[\xi_k(\hat{x}^{\dagger}_k\hat{m}+\hat{x}_k\hat{m}^{\dagger})+\mu_k(\hat{x}_k\hat{m}
+\hat{x}^{\dagger}_k\hat{m}^{\dagger})]$ \cite{Xie021046,Chen043708, Nakajima363,Ai042116,Zheng559}, where $\hat{m}^\dagger(\hat{m})$ and $\hat{x}_k^\dagger(\hat{x}_k)$ correspond to the creation (annihilation) operators for the optical modes, while $\xi_k$ and $\mu_k$ respectively denote the coupling strengths of the rotating wave and non-rotating-wave interactions between optical modes and non-Markovian environments. Our results might also be extended to possible applications including manipulating photon transport and quantum communication protocols with non-Markovian effects.
\section*{ACKNOWLEDGMENTS}
H. Z. S. acknowledges National Natural Science Foundation of China under Grants
No. 12274064 and Scientific Research Project for Department of Education
of Jilin Province under Grant No. JJKH20241410KJ. C. S. acknowledges financial support from the China Scholarship Council, the Japanese Government (Monbukagakusho-MEXT) Scholarship (Grant No. 211501), the RIKEN Junior Research Associate Program, and the Hakubi Projects of RIKEN. Y. H. Z. acknowledges the National Natural Science Foundation of China (NSFC)(Grants Nos.~12374333).
\appendix
\section{The impulse response functions $\kappa_\nu(t)$ and correlation function $f_\nu(t)$ can be realized through the pseudomode theory
}\label{APP.1}

To exhibit the controllability of Lorentzian spectrum density in non-Markovian environments, we give an concrete example using Markovian pseudomodes method \cite{Jack0438032001,Barnett1997,Garraway22901997,Garraway46361997,Man0621042014,Man57632015,Mazzola0121042009,Pleasance0621052017}. We construct a system including
the Markovian environment, whose total Hamiltonian is  
\begin{align}
{{\hat H_{tot}}}={{\hat H}_S} + {{\hat H}_R} + {{\hat H}_I},\label{HIHS}
\end{align}
with
\begin{align}
{{\hat H}_S} =& \ \hbar {\omega _u}{{\hat u}^\dag }\hat u + \hbar {\omega _r}{{\hat r}^\dag }\hat r + \hbar {g_{ur}}(\hat u{{\hat r}^\dag } + {{\hat u}^\dag }\hat r),\label{xuanzhuanqianHS}\\
{{\hat H}_I} =& \ \hbar \sum\nolimits_k {{T_k}} (\hat l_k^\dag \hat r + {{\hat r}^\dag }{{\hat l}_k}),{{\hat H}_R} = \hbar \sum\nolimits_k {{\omega _k}\hat l_k^\dag {{\hat l}_k}},\label{xuanzhuanqianHI}
\end{align}
where $\hat H_{S}$ represents system Hamiltonian consisting of a cavity mode with eigenfrequency ${\omega _u}$ and a pseudomode
with eigenfrequency $\omega _r$. ${\hat u}$ represents cavity mode annihilation operator and $\hat r$ denotes pseudomode annihilation operator, which satisfy the bosonic commutation relations $[\hat u,  {\hat u}^\dag ] = 1$ and $[\hat r,  {\hat r}^\dag ] = 1$. The third term corresponds to the tunneling coupling (coupling strength ${g_{ur}}$) between cavity mode and pseudomode. Here, ${\hat H}_I$ is interaction Hamiltonian between pseudomode and Markovian environment
(annihilation operator $\hat l_k$) with coupling strength ${T_k} = \sqrt {\gamma /2\pi }$ and decay rate $\gamma$. Also, $\hat H_R$ denotes Hamiltonian of Markovian environment and the correlation function satisfies $[{\hat l_k}, \hat l_{k'}^\dag ] = {\delta _{kk'}}$. According to Eq.~(\ref{xuanzhuanqianHS}), the Heisenberg-Langevin equations with the Markovian approximation are ($\hbar = 1$)}\cite{Gardiner2000,Walls1994}
\begin{eqnarray}
\frac{d}{{dt}}\hat u &=&  - i{\omega _u}\hat u - i{g_{ur}}\hat r, \label{dota}\\
\frac{d}{{dt}}\hat r &=&  - i{g_{ur}}\hat u - \frac{\gamma }{2}\hat r - \sqrt \gamma  {{\hat l}_{in}}(t).
\label{dotb}
\end{eqnarray}
By solving Eq.~(\ref{dotb}), we obtain 
\begin{align}
\hat r(t) = & \ \hat r(0){e^{ - \frac{\gamma }{2}t}} - i{g_{ur}}\int_0^t {\hat r(\tau ){e^{ -  \frac{\gamma }{2}(t - \tau )}}d\tau } \nonumber\\&- \sqrt \gamma  \int_0^t {{{\hat l}_{in}}(\tau ){e^{ - \frac{\gamma }{2}(t - \tau )}}d\tau }
\end{align}
with ${\hat l_{{\rm{in}}}}(t){\rm{ = }}\sum\nolimits_k {{e^{ - i{\omega _k}t}}} {\hat l_k}/\sqrt {2\pi }$. 

Substituting $\hat r(t)$ into Eq.~(\ref{dota}), we get the integro-differential equation
\begin{equation}
\begin{aligned}
\frac{d}{{dt}}\hat u =  - i{\omega _u}\hat u - \int_0^t {\Theta(t - \tau )\hat z(\tau )d\tau }-\hat{U}(t),
\label{Heisenbergequationsofmotion3}
\end{aligned}
\end{equation}
where $\hat U(t) = i{g_{ur}}\hat r(0){e^{ -\frac{\gamma }{2}t}}-i {g_{ur}}
\sqrt \gamma  \int_0^t {{{\hat l}_{in}}(\tau ){e^{ - \frac{\gamma }{2}
(t - \tau )}}d\tau }$ represents the operator for the non-Markovian composite environment, which includes pseudomode plus its Markovian environment and $\Theta(t) = g_{ur}^2{e^{ -  \frac{\gamma }{2}t}} $ denotes correlation function.
The Lorentzian spectrum density $J(\omega )$ corresponding to the correlation function $\Theta(t) = g_{ur}^2{e^{ -  \frac{\gamma }{2}t}} \equiv \int {J(\omega )} {e^{ - i\omega t}}d\omega $ in Eq.~(\ref{Heisenbergequationsofmotion3}) is equal to Eq.~(\ref{J}),
where
\begin{equation}
\begin{aligned}
\lambda_{\nu}  = \gamma /2, \Gamma_{\nu}  = 4g_{ur}^2/\gamma,
\label{R29}
\end{aligned}
\end{equation}
which leads to
\begin{equation}
\begin{aligned}
\Theta(t-\tau) = \frac{1}{2}\Gamma_{\nu}\lambda_{\nu}e^{-\lambda_{\nu}(t-\tau)},
\label{R29}
\end{aligned}
\end{equation}
which is consistent with the definition of $f_{\nu}(t)$ in Eq.~(\ref{J}). With the operators expectation values defined by
$u\equiv\left\langle\hat{u}\right\rangle$, $r \equiv \langle \hat{r} \rangle$,
$l_k \equiv \langle \hat{l}_k \rangle$, $l_{in} \equiv \langle \hat{l}_{in} \rangle$, $U(t) \equiv \langle \hat{U}(t) \rangle$, $a_{in} \equiv \langle \hat{a}_{in} \rangle$, $K(t)\equiv \langle \hat{K}(t) \rangle$, $v\equiv\left\langle\hat{v}\right\rangle$ and developing the equality
\begin{equation}
\begin{aligned}
K(t) \equiv U(t),
\end{aligned}
\end{equation}
we can get
\begin{equation}
\begin{aligned}
l_{in}(t)&=\frac{\gamma}{2}v(t)+v^{\prime}(t),
\label{R1}
\end{aligned}
\end{equation}
\begin{equation}
\begin{aligned}
r=-\frac{i}{g_{ur}}\int_{-\infty}^{+\infty}L^{*}(-\tau)a_{in}(\tau)d\tau,
\label{R2}
\end{aligned}
\end{equation}
with
\begin{equation}
\begin{aligned}
v(t)=\frac{g_{ur}re^{-\frac{\gamma}{2} t}+iK(t)}{g_{ur}\sqrt{\gamma}},
\label{R3}
\end{aligned}
\end{equation}
\begin{equation}
\begin{aligned}
K(t)=\int_{-\infty}^{+\infty}L^{*}(t-\tau)a_{in}(\tau)d\tau.
\label{R4}
\end{aligned}
\end{equation}
The value of $r=\mathrm{Tr}[\hat{r}(0)\rho_r(0)]$ is determined by the initial state $\rho_r(0)$ of the pseudomode. If $r$ sets
\begin{equation}
\begin{aligned}
r=-\frac{\sqrt{\Gamma}\lambda}{g_{ur}}\int_{0}^{+\infty}e^{-\lambda\tau}a_{in}(\tau)d\tau.
\label{R5}
\end{aligned}
\end{equation}
Comparing Eq.~(\ref{R2}) and Eq.~(\ref{R5}), we can obtain
\begin{equation}
\begin{aligned}
L(t)=i\sqrt{\Gamma}\lambda e^{\lambda t}\theta(-t),
\label{R6}
\end{aligned}
\end{equation}
which corresponds to $\kappa_\nu(t)$ above Eqs.~(\ref{ga})-(\ref{gc}).
$\hat{u}$ can denote $\hat{a}$, $\hat{b}$, and $\hat{c}$ in Eqs.~(\ref{dotak})-(\ref{dotck}), which is reasonable according to Eq.~(\ref{R6}). Hence, we can conclude that if Eqs.~(\ref{R29}), (\ref{R1}), (\ref{R2}), and (\ref{R5}) are satisfied simultaneously, the equations derived using the Markovian pseudomode method agree exactly with Eqs.~(\ref{dotak})-(\ref{dotck}) with
non-Markovian conditions.

\end{document}